\pdfoutput=1
\documentclass[cernpreprint, texlive=2023, UKenglish, texmf, orcidlogo]{atlasdoc}
 
\usepackage{atlaspackage}
\usepackage{atlasbiblatex}
 
\usepackage{atlasphysics}
 
\graphicspath{{logos/}{figures/}}
 
\usepackage{ANA-HIGG-2019-10-PAPER-defs}
 
\usepackage{multirow}
\usepackage{adjustbox}
\usepackage{enumitem}
 
\usepackage{tikz}
\usepackage{tikz-3dplot}
\tikzset{>=latex} 
\usetikzlibrary{decorations.pathmorphing} 
 
\usetikzlibrary{positioning}
 
\usepackage{makecell}

\AtlasTitle{Measurement of the $CP$ properties of Higgs boson interactions with $\tau$-leptons with the ATLAS detector}

\AtlasJournalRef{Eur. Phys. J. C (2023) 83:563}
\AtlasDOI{10.1140/epjc/s10052-023-11583-y}
\PreprintIdNumber{CERN-EP-2022-244}
\journal{EPJC}
 
\AtlasAbstract{
A study of the charge conjugation and parity ($CP$) properties of the interaction between the Higgs boson and $\tau$-leptons is presented. The study is based on a measurement of $CP$-sensitive angular observables defined by the visible decay products of $\tau$-leptons produced in Higgs boson decays. 
The analysis uses 139 fb$^{-1}$ of proton–proton collision data recorded at a centre-of-mass energy of $\sqrt{s}= 13$~TeV with the ATLAS detector at the Large Hadron Collider. Contributions from $CP$-violating interactions between the Higgs boson and $\tau$-leptons are described by a single mixing angle parameter $\phi_{\tau}$
in the generalised Yukawa interaction.
Without constraining the $H\rightarrow\tau\tau$ signal strength to its expected value under the Standard Model hypothesis,
the mixing angle $\phi_{\tau}$ is measured to be $9^{\circ} \pm 16^{\circ}$, with an expected value of $0^{\circ} \pm 28^{\circ}$ at the 68\% confidence level.
The pure $CP$-odd hypothesis is disfavoured at a level of 3.4~standard deviations.
The results are compatible with the predictions for the Higgs boson in the Standard Model.
}
 
\author{The ATLAS Collaboration}
 
\AtlasRefCode{HIGG-2019-10}

\hypersetup{pdftitle={ATLAS document},pdfauthor={The ATLAS Collaboration}}
 
\begin{document}
 
\maketitle

\clearpage

% The next lines are included from the .//source/introduction.tex input file
\section{Introduction}
\label{sec:intro}
 
A detailed analysis of Higgs boson ($H$) decays into $\tau$-lepton pairs observed at the LHC~\cite{HIGG-2013-32, CMS-HIG-13-004, HIGG-2015-07}
allows a direct probe of the charge conjugation and parity ($CP$) properties of the Yukawa coupling of the Higgs boson to the $\tau$-lepton.
The Standard Model (SM) of particle physics predicts the Higgs boson to be a $CP$-even (scalar) particle.
The presence of a $CP$-odd (pseudoscalar) admixture has not yet been excluded, 
and any observed $CP$-odd contribution to the $H\tau\tau$ coupling properties would be a sign of physics beyond the SM.
 
Studies of the $CP$ properties of Higgs boson interactions with gauge bosons
performed by the ATLAS and CMS
experiments~\cite{HIGG-2015-06, HIGG-2016-22, HIGG-2016-21, CMS-HIG-14-035, CMS-HIG-18-002, CMS-HIG-17-034}
have shown no deviation from the SM predictions.
However, these measurements probe the bosonic couplings in which $CP$-odd contributions
enter only via higher-order operators
that are suppressed by powers of 1/$\Lambda^2$~\cite{efttop:2011prd},
where $\Lambda$ is the scale of the new physics in an effective field theory.
In contrast,
a $CP$-odd contribution to Yukawa couplings can be present at tree level. 
Recently, measurements of the $CP$ properties of the interaction between the Higgs boson and top quarks were performed
by the ATLAS~\cite{HIGG-2019-01} and CMS~\cite{CMS-HIG-19-013} Collaborations,
and excluded a pure $CP$-odd structure for the top-quark Yukawa
coupling at 3.9$\sigma$ and 3.2$\sigma$, respectively.
 
This paper presents a measurement of the $CP$ properties of the Higgs boson interaction with $\tau$-leptons.
The measurement is based on
$CP$-sensitive angular observables
defined using the visible $\tau$-lepton decay products.
Ideas about how to probe a $CP$-odd and $CP$-even admixture in the $\tau$-lepton Yukawa coupling 
in \Htautau decay were initially developed 
in the context of $e^+e^-$ colliders~\cite{DellAquila:1988sbm, Kramer:1993jn, Bower:2002zx, Desch:2003rw, Berge:2013jra}.
Originally, hadronic decays of the $\tau$-leptons into $\pi^{\pm} \nu$ and $\rho^{\pm} \nu$ were used, and observables
sensitive to the transverse spin correlations between the $\tau$-lepton decay products were constructed.
These methods, extended to $\ell^{\pm} (=e^{\pm},\mu^{\pm})\nu \nu$
and $a_1^{\pm} \nu$ decays and re-evaluated in the context
of $pp$ collisions at the LHC~\cite{Berge:2008wi, Berge:2008dr, Berge:2011ij, Berge:2014sra,Berge:2014wta, Berge:2015nua},
are adopted in this analysis.
Recently, a similar study was also performed by the CMS Collaboration~\cite{CMS-HIG-20-006}.
 
The general effective Yukawa interaction between the Higgs boson and $\tau$-leptons can be parameterised as in Refs.~\cite{Berge:2014sra, Berge:2015nua}:
\begin{equation*}
\mathcal{L}_{H\tau\tau} = - \frac{m_{\tau}}{\upsilon} \kappa_{\tau} (\cos\phitau\bar{\tau}\tau + \sin \phitau\bar{\tau}i\gamma_{5}\tau)H,
\end{equation*}
where $\upsilon = 246$~\GeV\ is the vacuum expectation value of the Higgs field,
$\kappa_{\tau}$ is the reduced Yukawa coupling strength,
and $\phi_{\tau}$ (where  $\phi_{\tau} \in [-90^{\circ}, 90^{\circ}]$)
is the $CP$-mixing angle that parameterises the relative contributions of the $CP$-even and $CP$-odd components to the $H\tau\tau$ coupling.
The SM $CP$-even hypothesis is realised for $\phi_{\tau} = 0^{\circ}$,
while the pure $CP$-odd scenario corresponds to $\phi_{\tau} = \pm 90^{\circ}$.
Other values of  $\phi_{\tau}$ represent admixtures of the two components and would indicate a $CP$-violating scenario.

\newcommand{\PolarimeterFootnote}{
A polarimeter vector is calculated from the matrix element of the $\tau$-lepton decay process and is usually expressed in terms of the momenta of the decay products. Its direction gives the most probable direction of the $\tau$ polarization vector, while its magnitude determines the efficiency of a given decay as a polarimeter~\cite{Jadach:1984iy}.
}
 
The $CP$-mixing angle  $\phi_{\tau}$ is encoded in the correlations between the transverse spin components of the $\tau$-leptons in the $H \to \tau \tau$ decays,
which are then reflected in the directions of the $\tau$-lepton decay products.
The signed acoplanarity angle \phistar between the $\tau$ decay planes
(described in Section~\ref{sec:strategy} and illustrated in Figure~\ref{fig:phistar_methods})
is sensitive to the transverse spin correlations
impacted by the $CP$-mixing angle of the Yukawa coupling.
Such correlations are usually calculated by contracting the polarimeter vectors
of the decayed $\tau$-leptons\footnote{\PolarimeterFootnote}
and the spin density matrix of the $\tau$-lepton-pair spin state, $R_{i,j}$, which depends on the $\tau$-lepton pair-production process~\cite{Kuhn:1982di, Jadach:1993hs, Davidson:2010rw}.
In the case of Higgs boson decays, the spin density matrix $R_{i,j}$ has only transverse components with respect to the $\tau$-lepton direction, and these are first-order trigonometric polynomials in the $2 \phitau$ angle.
Per-event sensitivity to $CP$-mixing depends on the $\tau$-lepton-pair decay modes and on the way in which the polarimeter vectors and decay planes are reconstructed from observable quantities.
The \phistar angle
is directly related to \phitau in the \Htautau differential decay rate and the relation has the form of a first-order trigonometric polynomial in $\cos(\phistar - 2 \phitau)$ at leading order~\cite{Kramer:1993jn,Berge:2014sra,Berge:2014wta}:
\begin{equation*}
\dif\mathrm{\Gamma}_{H\to\tautau} \approx 1 - b(E_{+})b(E_{-})\frac{\pi^2}{16}\cos(\phistar-2\phitau),
\label{eq:differential_decay_phistar}
\end{equation*}
where
$E_{\pm}$ are the energies of the charged decay particles in their respective $\tau$-lepton rest frames,
and
$b(E_{\pm})$ are the spectral functions~\cite{Tsai:1971vv} describing the spin analysing power of a given decay mode.
Different methods~\cite{Bower:2002zx, Desch:2003rw, Berge:2013jra,Berge:2008wi, Berge:2008dr, Berge:2011ij, Berge:2014sra,Berge:2014wta, Berge:2015nua} have been developed in an attempt to approximately reconstruct $\tau$-lepton decay planes.
The \phistar variable used in this analysis is constructed with various methods depending on the $\tau$-lepton decay modes, largely following
the strategy presented in Ref.~\cite{Berge:2015nua}.
 
The analysis is performed using 139~fb$^{-1}$ of $\sqrt{s} = 13$~\TeV\ proton--proton ($pp$) collision data
recorded from 2015 to 2018 with the ATLAS detector.
Two $\tau$-lepton-pair decay channels are considered in the analysis: the first with one leptonically (\taulep) and one hadronically decaying $\tau$-lepton (\tauhad),
referred to as the \tlhad channel,
and the second with two hadronically decaying $\tau$-leptons, referred to as the \thadhad channel.
The leptonic decay $\tau^{\pm}\rightarrow \ell^{\pm}\nu\nu$
includes decays to either an electron or a muon.
In the case of hadronic decay, the dominant \tauhad decay modes are considered:
single-pion decay $\pi^{\pm}\nu$, 
two-pion decay $\pi^{\pm}\pi^{0}\nu$ with an intermediate $\rho^{\pm}$, 
and three-pion decay $\pi^{\pm} 2\pi^{0}\nu$ 
and $3 \pi^{\pm}\nu$ with an intermediate $a_1^{\pm}$. 
A small fraction of events with $\tau$ decays to $K^{\pm}$ mesons is also included in the analysis.
The $\tau$-lepton decay modes used in the analysis are summarised in \cref{tab:TauDecayModes} with their branching fractions~\cite{Zyla:2020zbs} and the notation used throughout this paper.
The \tauhad decay modes are labelled by YpXn in accord with the number of charged (Y) and neutral (X) pions among the decay products.
The $\tau$-lepton-pair decay modes considered in this analysis account for 68\% of all possible $\tau$ pair decays.

\begin{table}[t]
\centering
\caption{Notation for the dominant leptonic and hadronic $\tau$-lepton decay modes used and their branching fractions.
The symbol \enquote{$\ell^{\pm}$} stands for $e^{\pm}$ or $\mu^{\pm}$, and \enquote{$h^{\pm}$} includes $\pi^{\pm}$ and $K^{\pm}$.
The parentheses show the hadronic decays involving $\pi^{\pm}$ and their corresponding branching fractions.
}
\begin{tabular}{c|c|c}
\toprule
Notation &  Decay mode & Branching fraction\\
\midrule
$\ell$ &   $\ell^{\pm}\bar{\nu}\nu$ &   35.2\% \phantom{ (00.0\%)} \\ 
1p0n  & $h^{\pm}\nu$  ($\pi^{\pm}\nu$) & 11.5\%  (10.8\%) \\
1p1n & $h^{\pm} \pi^{0}\nu$ ($\pi^{\pm} \pi^{0}\nu$) & 25.9\%  (25.5\%)\\
1pXn & $h^{\pm} \geq 2\pi^{0}\nu$ ($\pi^{\pm} 2\pi^{0}\nu$)  & 10.8\% \phantom{0}(9.3\%) \\
3p0n & $3h^{\pm}\nu$ ($3\pi^{\pm}\nu$) & \phantom{0}9.8\% \phantom{0}(9.0\%) \\
\bottomrule
\end{tabular}
\label{tab:TauDecayModes}
\end{table}

This paper is structured as follows.
In \Cref{sec:detector} the ATLAS detector is briefly described.
The methodology and observables used in the analysis are discussed in \Cref{sec:strategy}.
\Cref{sec:sample} gives a summary of the data and simulated event samples.
\Cref{sec:object_and_selection} describes the object reconstruction and event selection, and defines
the signal and control regions.
\Cref{sec:systematics} details the experimental and theoretical systematic uncertainties.
The fit model and statistical analysis strategy are explained in \Cref{sec:statistical_analysis}.
\Cref{sec:results} presents the measurement results.
\Cref{sec:conclusion} concludes the paper.

% End of text imported from the .//source/introduction.tex input file

% The next lines are included from the .//source/detector.tex input file
\section{ATLAS detector}
\label{sec:detector}
 
\newcommand{\AtlasCoordFootnote}{
ATLAS uses a right-handed coordinate system with its origin at the nominal interaction point 
in the centre of the detector and the \(z\)-axis along the beam pipe.
The \(x\)-axis points from the interaction point to the centre of the LHC ring,
and the \(y\)-axis points upwards.
Cylindrical coordinates \((r,\phi)\) are used in the transverse plane,
\(\phi\) being the azimuthal angle around the \(z\)-axis.
The pseudorapidity is defined in terms of the polar angle \(\theta\) as \(\eta = -\ln \tan(\theta/2)\).
Angular distance is measured in units of \(\Delta R \equiv \sqrt{(\Delta\eta)^{2} + (\Delta\phi)^{2}}\).}

The ATLAS detector~\cite{PERF-2007-01} at the LHC covers nearly the entire solid angle around the collision point.\footnote{\AtlasCoordFootnote}
It consists of an inner tracking detector surrounded by a thin superconducting solenoid, electromagnetic and hadron calorimeters,
and a muon spectrometer incorporating three large superconducting air-core toroidal magnets.
 
The inner-detector system (ID) is immersed in a \SI{2}{\tesla} axial magnetic field
and provides charged-particle tracking in the range \(|\eta| < 2.5\).
The high-granularity silicon pixel detector covers the vertex region and typically provides four measurements per track,
the first hit normally being in the insertable B-layer~\cite{ATLAS-TDR-19,PIX-2018-001} installed before Run~2.
It is followed by the silicon microstrip tracker, which usually provides eight measurements per track.
These silicon detectors are complemented by the transition radiation tracker (TRT),
which enables radially extended track reconstruction up to \(|\eta| = 2.0\).
The TRT also provides electron identification information
based on the fraction of hits (typically 30 in total) above a higher energy-deposit threshold corresponding to transition radiation.
 
The calorimeter system covers the pseudorapidity range \(|\eta| < 4.9\).
Within the region \(|\eta|< 3.2\), electromagnetic calorimetry is provided by barrel and
endcap high-granularity lead/liquid-argon (LAr) calorimeters,
with an additional thin LAr presampler covering \(|\eta| < 1.8\)
to correct for energy loss in material upstream of the calorimeters.
Hadron calorimetry is provided by the steel/scintillator-tile calorimeter,
segmented into three barrel structures within \(|\eta| < 1.7\), and two copper/LAr hadron endcap calorimeters.
The solid angle coverage is completed with forward copper/LAr and tungsten/LAr calorimeter modules
optimised for electromagnetic and hadronic energy measurements respectively.
 
The muon spectrometer comprises separate trigger and
high-precision tracking chambers measuring the deflection of muons in a magnetic field generated by the superconducting air-core toroidal magnets.
The field integral of the toroids ranges between \num{2.0} and \SI{6.0}{\tesla\metre}
across most of the detector.
Three sets of precision chambers cover the region \(|\eta| < 2.7\) with multiple layers of monitored drift tubes,
complemented by cathode-strip chambers in the forward region, where the background is highest.
The muon trigger system covers the range \(|\eta| < 2.4\) with resistive-plate chambers in the barrel and thin-gap chambers in the endcap regions.
 
Interesting events are selected by the first-level trigger system implemented in custom hardware,
followed by selections made by algorithms implemented in software in the high-level trigger~\cite{TRIG-2016-01}.
The first-level trigger accepts events from the \SI{40}{\MHz} bunch crossings at a rate below \SI{100}{\kHz},
which the high-level trigger reduces in order to record events to disk at about \SI{1}{\kHz}.
 
An extensive software suite~\cite{ATL-SOFT-PUB-2021-001} is used in data simulation, in the reconstruction and analysis of real
and simulated data, in detector operations, and in the trigger and data acquisition systems of the experiment.

% End of text imported from the .//source/detector.tex input file

% The next lines are included from the .//source/strategy.tex input file
\section{Analysis strategy}
\label{sec:strategy}

\begin{figure}[htbp]
\centering
\subfloat[
$H\rightarrow\tau^+\tau^-\to\pi^+\pi^-+2\nu$ 
]{
\includegraphics[]{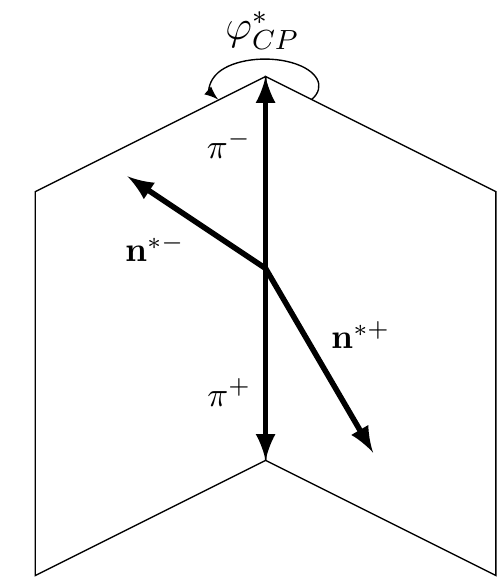}
\label{fig:ip_method}
}
\hfill
\subfloat[
$H\rightarrow\tau^+\tau^-\rightarrow\pi^+\pi^0\nu\pi^-\pi^0\nu$ 
]{
\includegraphics[]{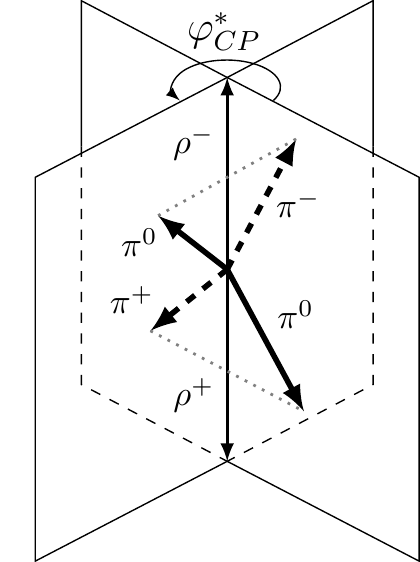}
\label{fig:rho_method}
}
\hfill
\subfloat[
$H\rightarrow\tau^+\tau^-\rightarrow\pi^+\pi^0\nu\pi^-\nu$ 
]{
\includegraphics[]{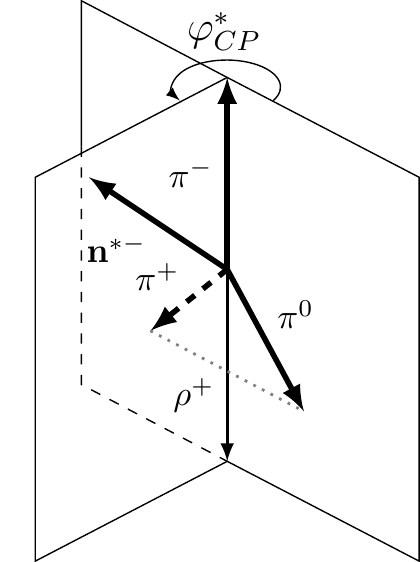}
\label{fig:ip_rho_method}
}
 
\caption{Illustration of the $\tau$-lepton decay planes for constructing the \phistar observable
in
\protect\subref{fig:ip_method} $H\rightarrow\tau^{+}\tau^{-}\to\pi^{+}\pi^{-}+2\nu$ decay using the impact parameter method,
\protect\subref{fig:rho_method} $H\rightarrow\tau^{+}\tau^{-}\rightarrow\pi^{+}\pi^{0}\nu\pi^{-}\pi^{0}\nu$ using the $\rho$-decay plane method,
and
\protect\subref{fig:ip_rho_method} $H\rightarrow\tau^{+}\tau^{-}\rightarrow\pi^{+}\pi^{0}\nu\pi^{-}\nu$ using the combined impact parameter and $\rho$-decay plane method.
The decay planes are spanned by the spatial momentum vector of the charged decay particle of the $\tau$-lepton ($\pi^{\pm}$) and either its impact parameter $\mathbf{n}^{*\pm}$ or the spatial momentum vector of the neutral decay particle of the $\tau$-lepton (\pizero).
}
\label{fig:phistar_methods}
\end{figure}

A $CP$-sensitive observable \phistar is built with different methods depending on the $\tau$-lepton decay modes.
In general, \phistar is the signed acoplanarity angle between the $\tau$-lepton decay planes.
Each $\tau$ decay plane is constructed from
the spatial momentum vector of a charged decay particle
and either its impact parameter (impact parameter method) or
the spatial momentum vectors of other visible $\tau$-lepton decay particles ($\rho$-decay plane and $a_1$ methods).
All vectors are boosted to the zero-momentum frame (ZMF) of the visible $\tau$-lepton-pair decay particles.
Figures~\ref{fig:phistar_methods}\subref{fig:ip_method}--\ref{fig:phistar_methods}\subref{fig:ip_rho_method}
illustrate the methods used to construct the \phistar observable in
$H\to\tau^{+}\tau^{-}\to\pi^{+}\pi^{-}+2\nu$,
$H\rightarrow\tau^{+}\tau^{-}\rightarrow\pi^{+}\pi^{0}\nu\pi^{-}\pi^{0}\nu$ 
and
$H\rightarrow\tau^{+}\tau^{-}\rightarrow\pi^{+}\pi^{0}\nu\pi^{-}\nu$ 
decays, respectively.
The visible $\tau$-lepton-pair ZMF (indicated by $^{*}$)
is used to approximate the Higgs boson rest frame,
which is not accessible due to the presence of undetected neutrinos in the $\tau$-lepton decays.

\begin{figure}[htbp]
\centering
\includegraphics[width=0.5\textwidth]{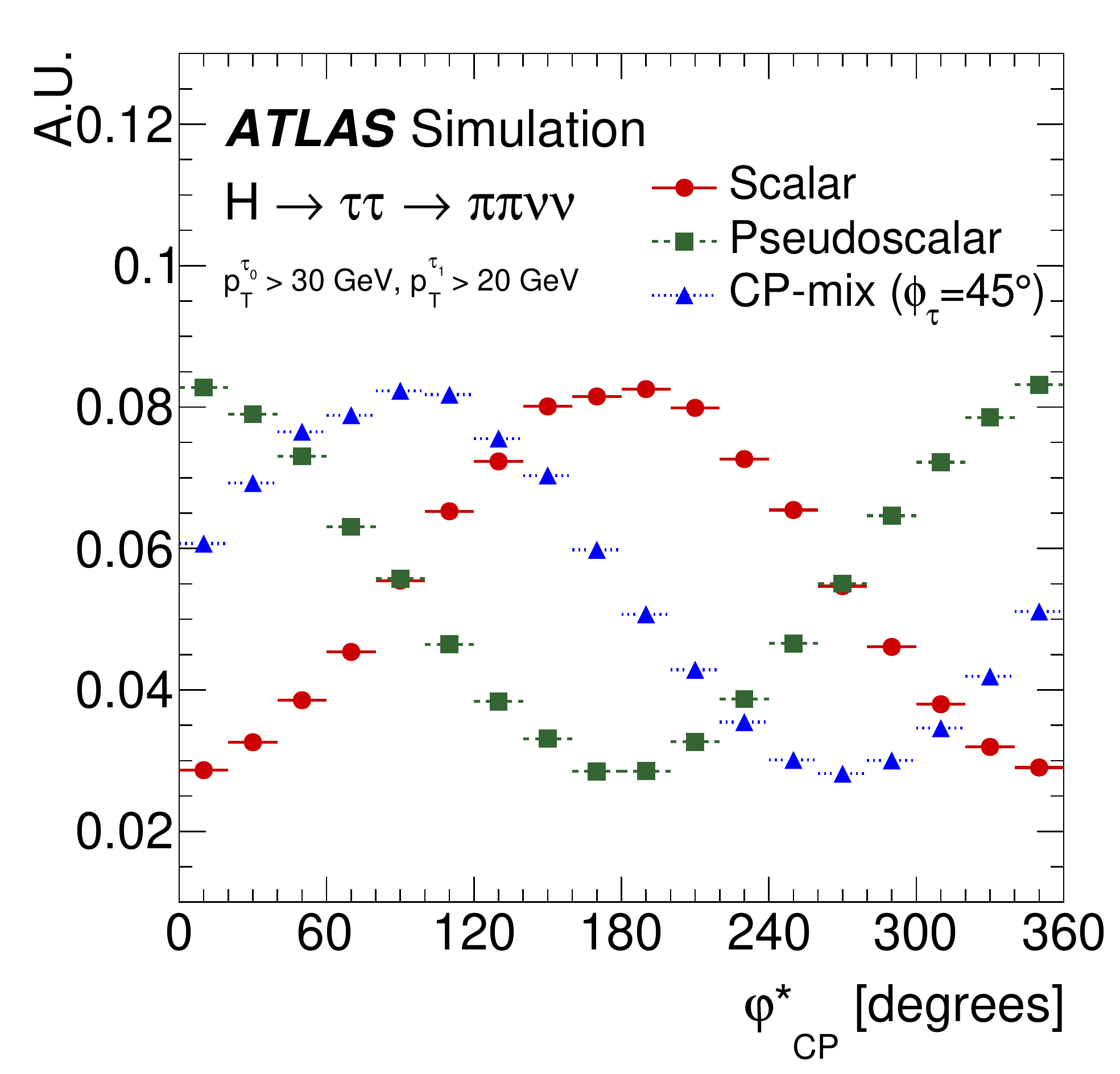}
\\
\caption{Normalised \phistar distributions
in simulated $H\to\tau^{+}\tau^{-}\to\pi^{+}\pi^{-}+2\nu$ events at the generator level for different $CP$ hypotheses.
The predictions for a pure $CP$-even SM Higgs boson (scalar, red circle), a pure $CP$-odd hypothesis (pseudoscalar, green square), and $CP$-mix hypothesis (\phitau = 45$^{\circ}$, blue triangle) are shown.
The transverse momentum of the simulated $\tau$ leptons is required to be larger than 30~\GeV\ (20~\GeV) for the leading (sub-leading) $\tau$ lepton during the event generation.
}
\label{fig:phistar_truth_ipip_VBFH_hh}
\end{figure}

Figure~\ref{fig:phistar_truth_ipip_VBFH_hh} shows the normalised distribution of \phistar
for simulated $H\rightarrow\tau^{+}\tau^{-}\rightarrow\pi^{+}\pi^{-}+2\nu$ events at the generator level.
The distribution peaks at \phistar = 180$^{\circ}$ for a $CP$-even (e.g.\ SM) Higgs boson,
whereas for the case of a pure $CP$-odd Higgs boson, the distribution peaks at \phistar = 0$^{\circ}$ and 360$^{\circ}$.
The phase difference between the \phistar distributions for two different mixing scenarios is twice their \phitau difference.
 
The $\tau$-lepton-pair decay combinations used in the analysis and the respective methods for constructing the \phistar observable are summarised in Table~\ref{tab:channels_methods}.
The corresponding fraction of events relative to the total from all possible di-$\tau$ decay combinations
is calculated from the single-$\tau$-lepton decay mode branching fractions in Table~\ref{tab:TauDecayModes}.
Other decay combinations are not considered in this analysis because their respective \phistar observables perform relatively poorly in discriminating between different $CP$ scenarios.

\begin{table}[t]
\centering
\caption{Decay mode combinations of the $\tau$-lepton pair and the corresponding methods to construct the \phistar observable used in this analysis.
The fraction of events for each decay mode combination relative to the total from all di-$\tau$ decay combinations (last column)
is calculated using the $\tau$-lepton decay mode branching fractions in Table~\ref{tab:TauDecayModes}.
}
\begin{tabular}{cccc}
\toprule
Decay channel & Decay mode combination & Method & Fraction in all $\tau$-lepton-pair decays \\
\midrule
\multirow{3}{*}{\tlhad} & $\ell$--1p0n & IP & 8.1\% \\
& $\ell$--1p1n & IP--$\rho$ & 18.3\%\phantom{0} \\
& $\ell$--1pXn & IP--$\rho$ & 7.6\% \\
& $\ell$--3p0n & IP--$a_{1}$ & 6.9\% \\
\midrule
\multirow{6}{*}{\thadhad} & 1p0n--1p0n & IP & 1.3\% \\
& 1p0n--1p1n & IP--$\rho$ & 6.0\% \\
& 1p1n--1p1n & $\rho$ & 6.7\% \\
& 1p0n--1pXn & IP--$\rho$ & 2.5\% \\
& 1p1n--1pXn & $\rho$ & 5.6\% \\
& 1p1n--3p0n & $\rho$--$a_{1}$ & 5.1\% \\
\bottomrule
\end{tabular}
\label{tab:channels_methods}
\end{table}

\subsection{Impact parameter (IP) method}
\label{sec:ip_method}

The IP method is applied to $\tau$-lepton decays with only one charged particle in the final state, specifically the direct hadronic decay
$\tau^{\pm}\rightarrow\pi^{\pm}\nu$ 
or leptonic decays $\tau^{\pm}\rightarrow \ell^{\pm}\nu\nu$. 
This refers to the 1p0n--1p0n and $\ell$--1p0n decay mode combinations.
In this case, the $\tau$-lepton decay plane is formed from the spatial momentum vector $\mathbf{q^{\pm}}$ of the charged particle ($\pi^{\pm}$, $\ell^{\pm}$)
and the three-dimensional (3D) impact parameter vector $\mathbf{n^{\pm}}$ of the charged particle,
defined as the directional distance of closest approach of the charged particle's track to the reconstructed primary vertex (PV) of the event.
The four-vectors of the track momentum $q^{\pm}_{\mu}$ and the impact parameter $n^{\pm}_{\mu}$ = (0, $\mathbf{n^{\pm}}$), initially defined and measured in the laboratory frame,
are boosted to the rest frame of the two charged decay particles (the visible di-$\tau$ ZMF, denoted by $^{*}$).
The boosted and normalised impact parameter vector $\mathbf{\hat{n}^{*\pm}}$ is then decomposed
into components which are parallel and transverse
($\mathbf{\hat{n}_{\perp}^{*\pm}}$) to the direction of the associated
normalised spatial momentum vector $\mathbf{\hat{q}^{*\pm}}$. Using these
vectors, an angle $\varphi^{*}$
and a \emph{CP}-odd
triple correlation $\mathcal{O}^{*}_{CP}$ are defined as
\begin{equation*}
\varphi^{*} = \arccos (\mathbf{\hat{n}}^{*+}_{\perp} \cdot \mathbf{\hat{n}}^{*-}_{\perp})
\quad \textrm{and} \quad
\mathcal{O}^{*}_{CP} = \mathbf{\hat{q}}^{*-} \cdot (\mathbf{\hat{n}}^{*+}_{\perp} \times \mathbf{\hat{n}}^{*-}_{\perp}),
\end{equation*}
and both
are incorporated in a single observable \phistar{} ($0 \leq
\phistar \leq 360^{\circ}$) defined by
\begin{equation}
\phistar =
\begin{cases}
\hfil \varphi^{*} & \text{if } \mathcal{O}^{*}_{CP} \geq 0 \\
360^{\circ} - \varphi^{*} & \text{if } \mathcal{O}^{*}_{CP} < 0.
\end{cases}
\label{eq:phistar_ip}
\end{equation}

In the case of leptonic decay, due to a different sign in the spectral function
for the leptonic $\tau$ decays~\cite{Tsai:1971vv,Berge:2011ij},
an additional shift by $180^{\circ}$ is applied to synchronise the phase in \phistar with the other decays.
 
\subsection{$\rho$-decay plane ($\rho$) method}
\label{sec:rho_method}
 
The $\rho$ method is applied to construct \phistar in events with 1p1n--1p1n or 1p1n--1pXn decay mode combinations.
In the case of consecutive decays $\tau^{\pm}\rightarrow\rho^{\pm} \nu$, $\rho^{\pm}\rightarrow\pi^{\pm}\pi^{0}$, 
the $\tau$-lepton decay plane
can be formed from the spatial momentum vectors of the charged pion ($\mathbf{q}^{\pm}$) and neutral pion ($\mathbf{q}^{0\pm}$).
The four-momentum vectors of the $\pi^{\pm}$ and $\pi^{0}$ are boosted to the rest frame of the $\rho$-meson pair (the visible $\tau$-lepton-pair ZMF).
The angle $\varphi^{*}$ and triple correlation $\mathcal{O}^{*}_{CP}$ are then defined in the same way
as in the IP method using the unit spatial vectors,
but replacing the impact parameter component with the neutral-pion vector,
\begin{equation*}
\varphi^{*} = \arccos (\mathbf{\hat{q}}^{*0+}_{\perp} \cdot \mathbf{\hat{q}}^{*0-}_{\perp})
\quad \textrm{and} \quad
\mathcal{O}^{*}_{CP} = \mathbf{\hat{q}}^{*-} \cdot (\mathbf{\hat{q}}^{*0+}_{\perp} \times \mathbf{\hat{q}}^{*0-}_{\perp}),
\end{equation*}
where $\mathbf{\hat{q}}^{*0+}_{\perp}$ and $\mathbf{\hat{q}}^{*0-}_{\perp}$ are the normalised vectors transverse to the direction of the associated charged pion for each neutral pion.
A signed observable $\varphi^{*\prime}$ is defined similarly to
\cref{eq:phistar_ip},
\begin{equation*}
\varphi^{*\prime} =
\begin{cases}
\hfil \varphi^{*} & \text{if } \mathcal{O}^{*}_{CP} \geq 0 \\
360^{\circ} - \varphi^{*} & \text{if } \mathcal{O}^{*}_{CP} < 0.
\end{cases}
\label{eq:phistar_prime_rhorho}
\end{equation*}
An additional requirement that depends on the sign of the product of $\tau$-lepton
spin-analysing functions
\(
y_{\pm}^{\rho} = (E_{\pi^{\pm}} - E_{\pi^{0}})/(E_{\pi^{\pm}} + E_{\pi^{0}}),
\)
where $E_{\pi^{\pm,0}}$ is the pion energy in the laboratory frame, is needed to define the observable
\phistar sensitive to the $CP$-mixing angle as
 
\begin{equation}
\phistar =
\begin{cases}
\hfil \varphi^{*\prime} & \text{if } y_{+}^{\rho}y_{-}^{\rho} \geq 0 \\
\varphi^{*\prime} + 180^{\circ} & \text{if } y_{+}^{\rho}y_{-}^{\rho} < 0.
\end{cases}
\label{eq:phistar_rhorho}
\end{equation}
 
In the case of 1pXn decays (e.g.\ $\tau^{\pm}\rightarrow a_{1}^{\pm} \nu\rightarrow\pi^{\pm}2\pi^{0}\nu$),
the sum of the four-momenta of all neutral pions is taken as the neutral component in the $\rho$ method.

\subsection{Combined IP and $\rho$ (IP--$\rho$) method}
\label{sec:ip_rho_method}
 
For events with the combinations 1p0n--1p1n, 1p0n--1pXn, $\ell$--1p1n, and $\ell$--1pXn,
the IP method and the $\rho$ method are combined to compute the
\phistar angle.
In the case of $H\rightarrow\tau\tau\rightarrow\pi^{\mp}\nu\rho^{\pm}\nu$, $\rho^{\pm}\rightarrow\pi^{\pm}\pi^{0}$ (1p0n--1p1n) events,
\phistar is defined in the visible $\tau$-lepton-pair ZMF by using the $\pi\rho$ rest frame.
One of the decay planes is defined using the IP method and the other with the $\rho$ method.
The quantities $\varphi^{*}$, $\mathcal{O}^{*}_{CP}$ and $\varphi^{*\prime}$ are calculated in a way analogous to that in Section~\ref{sec:rho_method}, but with the one-neutral-pion component replaced with the impact parameter vector defined in Section~\ref{sec:ip_method}.
The \phistar observable is defined as
\begin{equation}
\phistar =
\begin{cases}
\hfil \varphi^{*\prime} & \text{if } y^{\rho} \geq 0 \\
\varphi^{*\prime} + 180^{\circ} & \text{if } y^{\rho} < 0,
\end{cases}
\label{eq:phistar_iprho}
\end{equation}
with a phase shift of $180^{\circ}$ depending on the sign of $y^{\rho}$.

\subsection{$a_{1}$-decay ($a_{1}$) method}
\label{sec:a1_method}
 
The $a_{1}$ method is an extension of the $\rho$ method discussed in Section~\ref{sec:rho_method}, and is used for
$\tau^{\pm}\rightarrow a_{1}^{\pm}\nu$,
$a_{1}^{\pm}\rightarrow\pi^{\pm}\pi^{+}\pi^{-}$.
The $\tau$-lepton decay plane is defined by the charged pion ($\pi^{\pm}_1$) with the highest transverse momentum and the vector sum of the other two $\pi$ momenta. 
The observable $y_{\pm}^{\rho}$ used in the $\rho$ method is modified to take the effect of the $\pi$ masses into account, and is defined by $y_{\pm}^{a_{1}}$ adopting the convention in Ref.~\cite{1608.02609} as
\begin{equation*}
y_{\pm}^{a_{1}}=\frac{E_{2 \pi}-E_{\pi^{\pm}_1}}{E_{2 \pi} + E_{\pi^{\pm}_1}} - \frac{m_{3 \pi}^2-m_{\pi^{\pm}_1}^{2}+m_{2 \pi}^{2}}{2m^2_{3 \pi}},
\end{equation*}
 
where $m_{3\pi}$ is the invariant mass of the three charged pions from the $a_{1}^{\pm}$ decay,
and $m_{2\pi}$ ($E_{2\pi}$) refers to the invariant mass (energy) of the system of the two $\pi$ in the $\tau$ decay that do not have the highest transverse momentum.
 
Similarly to Section~\ref{sec:ip_rho_method}, the $a_1$ method is combined with the IP method for the $\ell$--3p0n events and
with the $\rho$ method for the 1p1n--3p0n events in the \tlhad and \thadhad channels, respectively.
In the $\ell$--3p0n case, \phistar is defined similarly to
Eq.~(\ref{eq:phistar_iprho}),
but with $y^{\rho}$ replaced by $y^{a_{1}}$.
For 1p1n--3p0n, the \phistar computation is analogous to
Eq.~(\ref{eq:phistar_rhorho}),
except that the product $y_{\pm}^{\rho}y_{\mp}^{a_{1}}$ is used to determine whether the $180^{\circ}$ shift is applied.

% End of text imported from the .//source/strategy.tex input file

% The next lines are included from the .//source/sample.tex input file
\section{Data and simulated event samples}
\label{sec:sample}

This analysis uses 139~fb$^{-1}$ of $\sqrt{s} = 13$~\TeV\ $pp$ collision data recorded by ATLAS with good operating conditions~\cite{DAPR-2018-01} in 2015--2018.
The data were collected using single-lepton or di-hadronic $\tau$ triggers~\cite{TRIG-2018-05,TRIG-2018-01,TRIG-2019-03,TRIG-2019-02}.
Events used in the \tlhad channel
were accepted by single-lepton triggers with $\pT$ thresholds of
24~\GeV\ (26~\GeV) and 20~\GeV\ (26~\GeV)
for electron and muon candidates, respectively, in the 2015 (2016--2018) dataset.
Events used in the \thadhad analysis channel were accepted by di-hadronic $\tau$-lepton triggers,
with a \pT threshold of 35~\GeV\ for the leading $\tau$ candidate and 25~\GeV\ for the sub-leading $\tau$ candidate.
Due to rising instantaneous luminosity, the di-hadronic $\tau$ triggers used
in the 2016--2018 data-taking period required an additional first-level triggered jet with $\pT > 25$~\GeV\ and $|\eta| <$ 3.2.

The analysis considers the four main Higgs boson production processes at the LHC:
gluon--gluon fusion (ggF),
vector-boson fusion (VBF),
associated production with a vector boson ($VH$),
and associated production with top-quark pair ($t \bar t H$).
The {\POWHEG} NNLOPS program~\cite{Hamilton:2013fea,Hamilton:2015nsa,Alioli:2010xd,Nason:2004rx,Frixione:2007vw} was used to model ggF Higgs boson  production with next-to-next-leading-order (NNLO) accuracy.
The VBF and $VH$ production processes were simulated with {\POWHEG} at NLO accuracy in QCD.
The production of $t \bar t H$ events was simulated using \POWHEGBOX[v2]~\cite{Frixione:2007nw,Nason:2004rx,Frixione:2007vw,Alioli:2010xd,Hartanto:2015uka} at NLO with the \NNPDF[3.0nlo]~\cite{Ball:2014uwa} PDF set.
In all signal events, the decays of $\tau$-leptons were modelled by \PYTHIA[8]~\cite{Sjostrand:2014zea} with no spin correlations.
The spin correlations are reintroduced with event-by-event weights modelling $CP$-mixing-dependent transverse spin correlations,
using the \verb!TauSpinner! package~\cite{Czyczula:2012ny,Przedzinski:2018ett,Przedzinski:2014pla}.
Background samples of \Vjets and diboson events were generated by \Sherpa[2.2.1]~\cite{Bothmann:2019yzt} (including $\tau$-lepton decays),
and $t \bar t$ and single-top samples were generated by \POWPY[8], with \PYTHIA also performing $\tau$-lepton decays.
The simulated event samples are shared with the cross-section measurement in the \Htautau decay channel~\cite{HIGG-2019-09}.

All samples of simulated events were passed through a full simulation of the ATLAS detector response~\cite{SOFT-2010-01} using \GEANT~\cite{Agostinelli:2002hh}.
The effects of multiple interactions in the same or neighbouring bunch crossings (pile-up) were modelled by overlaying each hard-scatter event with minimum-bias events, simulated using the soft QCD processes of \PYTHIA[8].
The simulated events were then weighted such that the distribution of the average number of interactions per bunch
crossing matches the one observed in data.

% End of text imported from the .//source/sample.tex input file

% The next lines are included from the .//source/selection.tex input file
\section{Event selection and background estimation}
\label{sec:object_and_selection}
 
The analysis of $CP$-mixing in the \Htautau channel requires the selection of signal events characterised by the presence of
isolated leptons, visible decay products in hadronic $\tau$ decays, jets and  missing transverse momentum.
Events with an isolated lepton (electron or muon) and a hadronic $\tau$ decay are used for the \tlhad analysis channel,
whereas events with two hadronic $\tau$ decays are used for the \thadhad channel.
Additional jets, the invariant mass and the transverse momenta of the $\tau$ lepton pairs
are used for event categorisation (Section~\ref{sec:selection}).
Various kinematic and identification requirements are used to suppress the background.
 
\subsection{Objects and decay mode reconstruction}
\label{sec:object}
 
The analysis shares the same object reconstruction and identification algorithms as the \Htautau cross-section measurement~\cite{HIGG-2019-09}.
Here the most important features specific to this $CP$ measurement are recalled:
 
\begin{itemize}
\item
Tracks measured in the ID are used to reconstruct interaction vertices~\cite{PERF-2015-01}, of which the one with the
highest sum of squared transverse momenta $\sum \pT^{2}$ of the associated tracks is selected as the primary vertex of the hard interaction.
 
\item
Electrons are reconstructed from topological clusters of energy deposits in the electromagnetic calorimeter which are
matched to a track reconstructed in the ID~\cite{EGAM-2018-01} and are required to satisfy \enquote{loose} isolation and
\enquote{medium} identification criteria.
 
\item
Muons are reconstructed from signals in the muon spectrometer matched with tracks inside the ID. They are required
to satisfy \enquote{loose} identification and \enquote{tight} isolation criteria based on track information~\cite{MUON-2018-03}.
 
\item
Jets are reconstructed using a particle-flow algorithm~\cite{PERF-2015-09}. It applies an anti-$k_t$ algorithm~\cite{Cacciari:2008gp,Fastjet}
with distance parameter $R = 0.4$ to noise-suppressed positive-energy topological clusters in the calorimeter
after subtracting energy deposits associated with primary-vertex-matched tracks, and including the track momenta
instead in the clustering, thereby improving the jet energy measurement.
Cleaning criteria are used to identify jets arising from non-collision backgrounds or noise in the
calorimeters~\cite{ATLAS-CONF-2015-029}. A dedicated jet-vertex-tagger algorithm~\cite{PERF-2014-03}
is used to remove jets that are identified as not being associated with the primary vertex of the hard interaction.
Similarly, a dedicated algorithm is used to suppress pile-up jets in the forward region~\cite{PERF-2016-06}.
 
\item
Hadronic $\tau$-lepton decays are reconstructed from the visible decay products (neutral pions or charged pions/kaons).
The reconstruction is seeded by jets reconstructed with an anti-$k_{t}$ algorithm
using calibrated topological clusters~\cite{PERF-2014-07} as inputs and a distance parameter of $R = 0.4$~\cite{ATLAS-CONF-2017-029}.
Reconstructed nearby tracks are matched to a hadronic $\tau$ candidate (\tauhadvis) if they exceed the value required for a multivariate discriminant determining the likelihood that the tracks are produced from the \tauhadvis decay.
A recurrent neural network identification algorithm~\cite{ATL-PHYS-PUB-2019-033} is trained to separate the \tauhadvis candidates from jets initiated by quarks or gluons, and
boosted decision tree (BDT) discriminants are used to help reject misidentified hadronic $\tau$ decays due to electrons.
In both cases, the \tauhadvis is required to satisfy \enquote{medium} criteria.
Reconstructed \tauhadvis objects are required to have one or three associated tracks,
and have $\pT > 20$~\GeV\ and $|\eta| < 2.47$, excluding the transition region between the barrel and endcap electromagnetic calorimeters ($1.37 < |\eta| < 1.52$).
 
\item
The missing transverse momentum (with magnitude \met) is reconstructed as the negative vector sum of the transverse momenta of leptons, \tauhadvis objects, jets, and a `soft-term'.
The `soft-term' is calculated as the vector sum of the \pT of tracks matched to the primary vertex but not associated with reconstructed leptons, \tauhadvis objects, or jets~\cite{PERF-2016-07}.
 
\item The invariant mass of the $\tau$-lepton-pair system, \MMC, is estimated with
an advanced likelihood-based algorithm named the Missing Mass Calculator (MMC)~\cite{Elagin:2010aw,HIGG-2019-09}.
 
\end{itemize}
 
The transverse (longitudinal) impact parameter \dzero ($z_{0}$) of a charged-particle track is defined as the distance in the transverse plane ($z$ direction) from the
primary vertex
to the track's point of closest approach in the transverse plane.
The impact parameters are used in the calculation of the \phistar observable
in the cases involving single-pion or leptonic decays (Section~\ref{sec:ip_method}),
as well as in one of the variables defining the signal regions (Section~\ref{sec:selection}).
 
An important addition in the $CP$ analysis is that it requires
the $\tau$-lepton reconstruction algorithms to categorise the hadronic $\tau$-lepton decays
by using the number of reconstructed tracks
and the capability to distinguish between single-$\pi^0$ and multi-$\pi^0$ clusters.
This allows the \tauhad candidates to be classified as being from the
{1p0n}, {1p1n}, {1pXn} or {3p0n} decay modes.
For this measurement it is crucial to
identify each decay mode with good efficiency and purity, and
to measure the momenta of the charged and neutral pions.
This is achieved through the \enquote{\tauhad Particle Flow} reconstruction algorithm~\cite{PERF-2014-06},
in which the four-momenta of the charged and neutral pions from the $\tau$ decay
are determined by combining measurements from the tracking detector and the calorimeter.
The decay mode classification is performed by counting the number of charged and neutral pions,
exploiting the kinematic properties of the $\tau$-lepton decay products
and the number of reconstructed photons by using BDTs.
Three BDTs are built respectively for the 1p0n vs 1p1n, 1p1n vs 1pXn and 3p0n vs 3pXn decay modes to improve the determination of the number of neutral pions in each case.
The efficiency (purity) of the classification is about 80\% (70\%--80\%) for the dominant decay modes 1p0n and 1p1n;
for the 3p0n decay mode the efficiency is over 90\%, with 90\% purity.

\subsection{Event selection}
\label{sec:selection}
 
Events are selected by requiring at least one hadronic $\tau$-lepton decay in the signature.
For events in the \tlhad channel, a \tauhadvis with $\pT > 30$~\GeV\ and an electron or muon with $\pT > 21.0$--27.3~\GeV\
are required, with the latter \pT cut depending on the data-taking period.
In the \thadhad channel, two \tauhadvis objects are required with \pT above 40 and 30~\GeV.
The \pT requirements are chosen to be above the \pT thresholds of the respective single-lepton and hadronic $\tau$ triggers to ensure trigger operation at the plateau efficiency.
The \tauhadvis and the lepton are required to match geometrically with their trigger counterparts in the \thadhad and \tlhad channels, respectively.
In both channels the two $\tau$-lepton candidates are required to have opposite electric charges.
The angular distances between the $\tau$-lepton candidates are required to be $\Delta R_{\tau\tau} < 2.5$ and $|\Delta \eta_{\tau\tau}| < 1.5$ ($0.6 < \Delta R_{\tau\tau} < 2.5$ and $|\Delta \eta_{\tau\tau}| < 1.5$) to reject non-resonant events in the \tlhad (\thadhad) channel.
All events require at least one jet with $\pT > 40$~\GeV.
Due to triggering conditions, this selection is tightened to $\pT > 70$~\GeV\ and $|\eta| < 3.2$ in the \thadhad channel.
The \met is required to be greater than 20~\GeV, and
requirements on the fraction of the $\tau$-lepton momentum carried by its visible decay products
are applied to further improve the invariant mass estimation.
In the \tlhad channel, the transverse mass \mT of the lepton-plus-\met system
is required to be less than \SI{70}{\GeV} in order to efficiently reject \Wjets processes.
These preselection criteria are the same as in the \Htautau cross-section measurement~\cite{HIGG-2019-09}.

Events are further categorised to target the vector-boson fusion (VBF category) and gluon--gluon fusion (Boost category) Higgs boson production modes.
The VBF category contains events with a second high-$\pT$ jet with $\pT^{j_2} > 30$~\GeV.
The two leading jets are required to satisfy the following kinematics:
$|\Delta \eta_{jj}| > 3.0$, $\eta_{j_1} \cdot \eta_{j_2} < 0$, and $m_{jj} > 400$~\GeV, with the
pseudorapidity values of \taulep or \tauhad lying between those of the two leading jets (yielding \enquote{central $\tau$-leptons}).
This kinematic selection is enhanced by splitting the VBF category into two regions, {VBF\_1} and {VBF\_0},
based on the output of a BDT-based VBF tagger~\cite{HIGG-2019-09},
with the VBF\_1 region having an enhanced fraction of VBF Higgs boson production events.
The Boost category targets events where the Higgs boson recoils against jets.
The \pT of the Higgs boson ($\pT^{\tau\tau}$) is computed as the magnitude of the vector sum of the transverse momenta of the $\tau$-leptons' visible decay products and the missing transverse momentum.
Events with $\pT^{\tau\tau}> 100$~\GeV\ that do not pass the VBF selection form the Boost category.
These events are further separated into
those with $\Delta R_{\tau\tau} < 1.5$ and $\pT^{\tau\tau} > 140$~\GeV\ (Boost\_1)
and those not passing these selections (Boost\_0).
The event categorisation is applied to both the \tlhad and \thadhad channels, as summarised in Table~\ref{tab:EventSelection_vbf_boost}.
 
For each category, a Higgs-enriched signal region is defined with $110 < \MMC < 150$~\GeV,
and a \Ztautau control region (CR) is defined with $60 < \MMC < 110$~\GeV,
with the other selection criteria for each category remaining the same.
Two additional control regions are defined using events with a
$\tau^{\pm}\rightarrow\rho^{\pm} \nu\rightarrow\pi^{\pm}\pi^{0}\nu$ decay
in the $60 < \MMC < 110$~\GeV\ range:
one region with $\ell$--1p1n events for the \tlhad channel, and the other with 1p1n--1p1n events for the \thadhad channel.
The events in these regions are defined so as to be statistically independent of the four \Ztautau control regions in Table~\ref{tab:EventSelection_vbf_boost}.
The use of the control regions is described in Section~\ref{sec:statistical_analysis}.

\begin{table}[t]
\caption{Summary of selection criteria for the VBF and Boost categories in this analysis. The criteria are common to the \tlhad and \thadhad decay channels.
}
\centering
\begin{tabular}{ m{0.22\textwidth}| m{0.22\textwidth}| m{0.22\textwidth}| m{0.22\textwidth}}
\toprule
\multicolumn{2}{c}{\textrm{VBF}}
& \multicolumn{2}{|c}{\textrm{Boost}} \\
\midrule
 
\multicolumn{2}{c|}{$\pT^{j_2} > 30$ GeV}
& \multicolumn{2}{c}{\multirow{5}{*}{\makecell{Not VBF\\ $\pT^{\tau\tau} > 100$ GeV}}} \\
\multicolumn{2}{c|}{$m_{jj} > 400$ GeV}
& \multicolumn{2}{c}{} \\
\multicolumn{2}{c|}{$|\Delta\eta_{jj}| > 3.0$}
& \multicolumn{2}{c}{} \\
\multicolumn{2}{c|}{$\eta_{j_1} \cdot \eta_{j_2} < 0$}
& \multicolumn{2}{c}{} \\
\multicolumn{2}{c|}{Central $\tau$-leptons}
& \multicolumn{2}{c}{} \\
\midrule
\midrule
\multicolumn{4}{c}{Signal region ($110 < \MMC < \SI{150}{\GeV}$)} \\
\midrule
\multicolumn{1}{c|}{VBF\_1}
& \multicolumn{1}{c|}{VBF\_0}
& \multicolumn{1}{c|}{Boost\_1}
& \multicolumn{1}{c}{Boost\_0} \\
\midrule
 
\multicolumn{1}{c|}{\multirow{2}{*}{BDT(VBF) $>$ 0}}
& \multicolumn{1}{c|}{\multirow{2}{*}{BDT(VBF) $<$ 0}}
& \multicolumn{1}{c|}{$\Delta R_{\tau\tau} < 1.5$ and}
& \multicolumn{1}{c}{$\Delta R_{\tau\tau} > 1.5$ or} \\
 
\multicolumn{1}{c|}{} & \multicolumn{1}{c|}{} & \multicolumn{1}{c|}{$\pT^{\tau\tau} > 140$ GeV} & \multicolumn{1}{c}{$\pT^{\tau\tau} < 140$ GeV} \\
 
\midrule
\midrule
\multicolumn{4}{c}{\Ztt control region ($60 < \MMC < \SI{110}{\GeV}$)} \\
\midrule
\multicolumn{1}{c|}{VBF\_1 Z CR}
& \multicolumn{1}{c|}{VBF\_0 Z CR}
& \multicolumn{1}{c|}{Boost\_1 Z CR}
& \multicolumn{1}{c}{Boost\_0 Z CR} \\
\bottomrule
\end{tabular}
\label{tab:EventSelection_vbf_boost}
\end{table}

\FloatBarrier

Depending on the decay mode combination, different additional selection criteria are applied to enhance the sensitivity of the \phistar construction method.
The \phistar from the IP method is less effective in discriminating between different \phitau values when the impact parameter vector has a magnitude smaller than, or similar to, its resolution.
The sensitivity of this method can be enhanced by using events with high significance of the track impact parameter in the transverse plane, \dsig,
defined as the transverse impact parameter $d_{0}$ divided by its resolution $\sigma(d_{0})$.
Events are therefore separated into two groups based on the value of $|\dsig|$ of the lepton in \taulep or the pion in \tauhad.
In the $\rho$ method, events with larger absolute values of the product $|y^{\rho}_{+}y^{\rho}_{-}|$ are more sensitive to \phitau.
This quantity is also used to separate the events into two groups.
In the case of the combined IP--$\rho$ method, the sensitivity of \phistar is enhanced by separating the events based on the values $|\dsig|$ and $|y^{\rho}|$.
For the IP--$a_1$ and $\rho$--$a_1$ methods the separation is based on $|y_{\pm}^{a_{1}}|$.
Details of the additional selection criteria are summarised in Tables~\ref{tab:AdditionalSelection_lh} and \ref{tab:AdditionalSelection_hh}.

Among the $\tau$-lepton pair decays used in this analysis,
the most sensitive (\enquote{dominant}) decay mode combinations are 1p0n--1p1n, 1p1n--1p1n, and 1p0n--1p0n in the \thadhad decay channel, and $\ell$--1p1n and $\ell$--1p0n in the \tlhad decay channel,
while the other combinations involving 1pXn and 3p0n are subdominant due to their
weaker spin analysing power
and smaller decay fractions.
The events are therefore divided into three groups, the \enquote{High}, \enquote{Medium} and \enquote{Low} signal regions, which characterise the different levels of sensitivity.
Events satisfying the additional selection criteria for the dominant and subdominant decay mode combinations  
define the High and Medium signal regions, respectively, while the rest are grouped into the Low signal region.
This allows the \phistar distributions from the decay mode combinations with similar sensitivity to $CP$-mixing to be merged
to increase the statistical precision of the distribution templates within each signal region,
allowing the use of finer binning in the \phistar distributions for the regions with better sensitivity.
The splitting of the signal regions summarised in Tables~\ref{tab:AdditionalSelection_lh} and \ref{tab:AdditionalSelection_hh} is applied in each of the VBF and Boost categories (VBF\_1, VBF\_0, Boost\_1 and Boost\_0).

\begin{table}[t]
\centering
\caption{Summary of additional selection criteria for the signal regions in the \tlhad channel.
}
\begin{tabular}{ cccc }
\toprule
Channel & Signal region & Decay mode combination & Selection criteria \\
\midrule
\multirow{11}{*}{\tlhad} & \multirow{4.5}{*}{High} & \multirow{2}{*}{$\ell$--1p0n} & $|\dsig(e)| > 2.5$ or $|\dsig(\mu)| > 2.0$ \\
& & & $|\dsig(\tau_{\textrm{1p0n}})| > 1.5$ \\
\cmidrule{3-4}
& & \multirow{2}{*}{$\ell$--1p1n} & $|\dsig(e)| > 2.5$ or $|\dsig(\mu)| > 2.0$  \\
& & & $|y^{\rho}(\tau_{\textrm{1p1n}})| > 0.1$  \\
\cmidrule{2-4}
& \multirow{4.5}{*}{Medium} & \multirow{2}{*}{$\ell$--1pXn} & $|\dsig(e)| > 2.5$ or $|\dsig(\mu)| > 2.0$ \\
& & & $|y^{\rho}(\tau_{\textrm{1pXn}})| > 0.1$ \\
\cmidrule{3-4}
& & \multirow{2}{*}{$\ell$--3p0n} & $|\dsig(e)| > 2.5$ or $|\dsig(\mu)| > 2.0$  \\
& & & $|y^{a_{1}}(\tau_{\textrm{3p0n}})| > 0.6$  \\
\cmidrule{2-4}
& Low & All above & Not satisfying selection criteria \\ 
\bottomrule
\end{tabular}
\label{tab:AdditionalSelection_lh}
\end{table}
 
\begin{table}[t]
\centering
\caption{Summary of additional selection criteria for the signal regions in the \thadhad channel.
}
\begin{tabular}{ cccc }
\toprule
Channel & Signal region & Decay mode combination & Selection criteria \\ 
\midrule
\multirow{16}{*}{\thadhad} & \multirow{7}{*}{High} & \multirow{2}{*}{1p0n--1p0n} & $|\dsig(\tau_1)| > 1.5$ \\
& & & $|\dsig(\tau_2)| > 1.5$ \\
\cmidrule{3-4}
& & \multirow{2}{*}{1p0n--1p1n} & $|\dsig(\tau_{\textrm{1p0n}})| > 1.5$ \\
& & & $|y^{\rho}(\tau_{\textrm{1p1n}})| > 0.1$ \\
\cmidrule{3-4}
& & \multirow{2}{*}{1p1n--1p1n} & \multirow{2}{*}{$|y^{\rho}(\tau_1)y^{\rho}(\tau_2)| > 0.2$} \\
& & & \\
\cmidrule{2-4}
& \multirow{7}{*}{Medium} & \multirow{2}{*}{1p0n--1pXn} & $|\dsig(\tau_{\textrm{1p0n}})| > 1.5$ \\
& & & $|y^{\rho}(\tau_{\textrm{1pXn}})| > 0.1$ \\
\cmidrule{3-4}
& & \multirow{2}{*}{1p1n--1pXn} & \multirow{2}{*}{$|y^{\rho}(\tau_{\textrm{1p1n}})y^{\rho}(\tau_{\textrm{1pXn}})| > 0.2$} \\
& & & \\
\cmidrule{3-4}
& & \multirow{2}{*}{1p1n--3p0n} & $|y^{\rho}(\tau_{\textrm{1p1n}})| > 0.1$ \\
& & & $|y^{a_{1}}(\tau_{\textrm{3p0n}})| > 0.6$ \\
\cmidrule{2-4}
& Low & All above & Not satisfying selection criteria \\ 
\bottomrule
\end{tabular}
\label{tab:AdditionalSelection_hh}
\end{table}

In this configuration, there are 12 signal regions in each of the \tlhad and \thadhad decay channels, leading to 24 signal regions in total. Each signal or control region is orthogonal to the others.
 
\FloatBarrier
 
\subsection{Background estimation}
\label{sec:backgrounds}
 
Expected SM processes other than the \Htautau signal are evaluated using a mixture of simulation and data-driven techniques.
Processes with true $\tau$-leptons and prompt light leptons are estimated through simulation.
Among those the \Zttjets process is the dominant one, and dedicated control regions with $60 < \MMC < 110 $~\GeV\ (Section~\ref{sec:selection}) are used to extract its normalisation (Section~\ref{sec:statistical_analysis}).

The second most significant background contribution arises from jets misidentified as hadronically decaying $\tau$-leptons (\tauhad), referred to as \enquote{misidentified-$\tau$ background}.
It is determined by a data-driven approach using fake factors, as described in detail in Ref.~\cite{HIGG-2019-09}.
The misidentified-$\tau$ background consists mostly of \Wjets, QCD multijet, and top-quark events in the \tlhad channel,
while in the \thadhad channel it originates mainly from multijet events.
The fake factors are calculated as the ratio of the number of events passing the $\tau$ identification requirements to the number failing them in dedicated QCD and \Wjets-enriched regions,
defined by inverting the lepton isolation or \mT requirement, respectively.
The fake factors are estimated in each hadronic $\tau$ decay mode considered in the analysis, separately for VBF and Boost region events.
To estimate both the shape and the normalisation of the misidentified-$\tau$ background,
distributions in the regions failing the $\tau$ identification are multiplied by the fake factors to correct for the different efficiencies to pass or fail the $\tau$ identification selection.
In the \enquote{failed $\tau$ identification} regions, events not corresponding to misidentified-$\tau$ background from jets are subtracted using simulated event samples.
For the \thadhad final state the fake-factor method is modified slightly:
the \tauhad objects are matched to their
trigger counterparts and the estimate covers processes with one or two jets misidentified as \tauhad.
The misidentified-$\tau$ background contribution is estimated for each di-$\tau$ decay combination considered (Section~\ref{sec:strategy}) in the preselection region for the \thadhad events.

Smaller background contributions are due to diboson, \Zlljets and $H\rightarrow WW^{*}$ processes in both decay channels, while \Wjets and top production also make small contributions in the \thadhad channel. These are estimated from simulation and are normalised to their theoretical expectations.

% End of text imported from the .//source/selection.tex input file

% The next lines are included from the .//source/systematics.tex input file
\section{Systematic uncertainties}
\label{sec:systematics}

Systematic uncertainties generally affect the yield in the signal and control regions as well as the shape of the \phistar distribution, on which the fit is performed.
They are grouped into three types: the experimental uncertainties, the theoretical uncertainties,
and the $\tau$-lepton decay reconstruction uncertainties.

The experimental uncertainties include those from the
trigger, reconstruction, identification and isolation requirements
of the final-state particle candidates
(electrons, muons, $\tau$-leptons, jets, \met, $b$-tagging),
as well as uncertainties from misidentified-$\tau$ background estimation and the luminosity measurement~\cite{ATLAS-CONF-2019-021,LUCID2}.

The uncertainties on the jet energy resolution and scale~\cite{JETM-2018-05} affect the acceptance of the signal and background contributions through their impact on \met, and thus \MMC~\cite{HIGG-2017-07}.
They also directly affect the jet selection for the event categorisation in the VBF and Boost regions~\cite{HIGG-2017-07}.
The jet energy scale uncertainty for central jets ($|\eta| <1.2$)
varies from
1\% for a wide range of jet \pT ($250~\GeV < \pT < 2000$~\GeV)
to 5\% for very low \pT jets ($\pT < 20$~\GeV),
and 3.5\% for very high \pT jets ($\pT > 2.5$~\TeV),
and forward jets exhibit uncertainties of similar size.
The relative jet energy resolution ranges from ($24 \pm 1.5$)\% to ($6 \pm 0.5$)\% for jets with \pT of 20~\GeV\ to 300~\GeV, respectively~\cite{JETM-2018-05}.
Other jet-related uncertainties have smaller impacts.

Uncertainties from the misidentified-$\tau$ background estimation arise from
sample statistics of the events used for estimating the fake factors,
subtraction of residual contributions from processes without misidentified-$\tau$, 
and uncertainties in the flavour composition taken from comparisons between the predicted and observed backgrounds in a dedicated validation region.
The uncertainties are assigned per $\tau$-lepton-pair decay combination in the \tlhad and \thadhad channels.
The total expected uncertainties in the yield of the misidentified-$\tau$ events in the signal regions are
typically 20\% (20\%--40\%) in the Boost (VBF) region in the \tlhad channel, and 15\%--30\% in the \thadhad channel.

Theoretical uncertainties are applied to \Zjets background and signal processes.
For the \Zjets background, uncertainties are considered for the
renormalisation
and factorisation
scales,
the resummation scale, 
the jet-to-parton matching scheme, 
the choice of value for the strong coupling constant \alphas,
and the choice of PDF.
Only distribution shapes and migrations between analysis regions are considered for these uncertainties since the absolute normalisation of \Ztautau is determined at the fit level.
For the signal, uncertainties are considered for
the QCD scale because of missing higher orders in the matrix element calculation,
for the parton shower and hadronisation model,
and for renormalisation and factorisation scales and the PDF.
These are applied to the production cross-sections for the ggF, VBF and $VH$ processes, and do not have large impacts on the measurement as they only affect the signal normalisation, which is also determined in the fit, and not the shape of the \phistar distribution.
 
The $\tau$-lepton decay reconstruction uncertainties primarily affect the shape of the \phistar distributions.
These concern the classification of the hadronic $\tau$-lepton decay modes
as well as measurement uncertainties of the track impact parameters, pion track momentum, and neutral pion momentum.
The uncertainties in the classification of the hadronic $\tau$-lepton decay modes
are derived from a $\tau$-lepton decay mode classification efficiency and correction factor measurement,
using a tag-and-probe analysis in the $\mu$--\tauhad final state performed on part of the Run~2 dataset.
These uncertainties include those in the decay mode reconstruction efficiencies and in the event migration between decay modes.
The size of the uncertainties ranges from 6\% to 20\% depending on the decay modes and their cross migrations.
The uncertainties affecting the impact parameters and track measurements include those in alignment effects 
and in impact parameter resolution, and they account for differences between data and simulation.
 
\begin{figure}[thp]
\centering
\subfloat[\tlhad ZCR with $\ell$--1p1n events]{\includegraphics[width=0.49\textwidth]{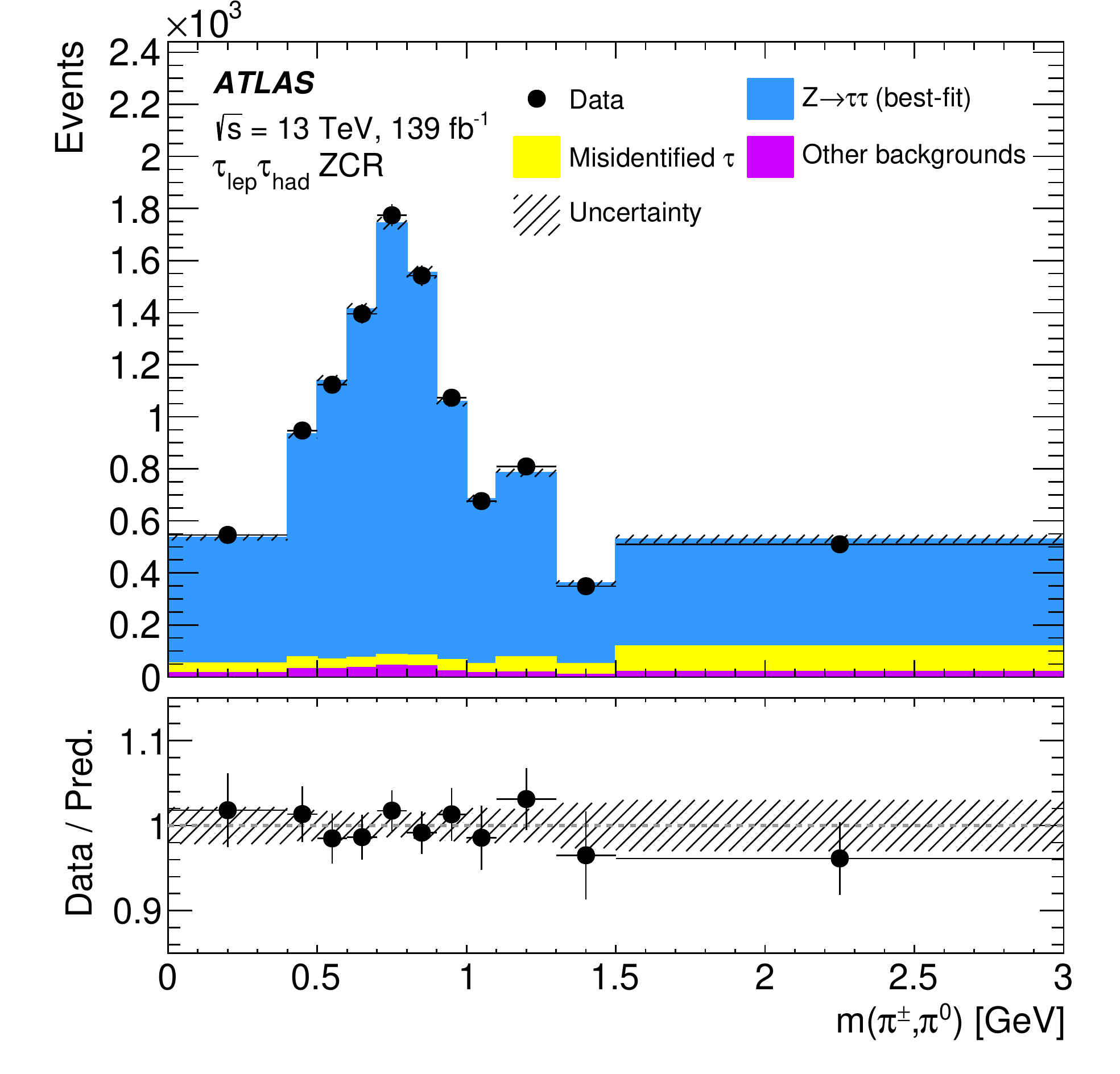}
\label{fig:ZttRhoCR_postfit_lh}
}
\subfloat[\thadhad ZCR with 1p1n--1p1n events]{\includegraphics[width=0.49\textwidth]{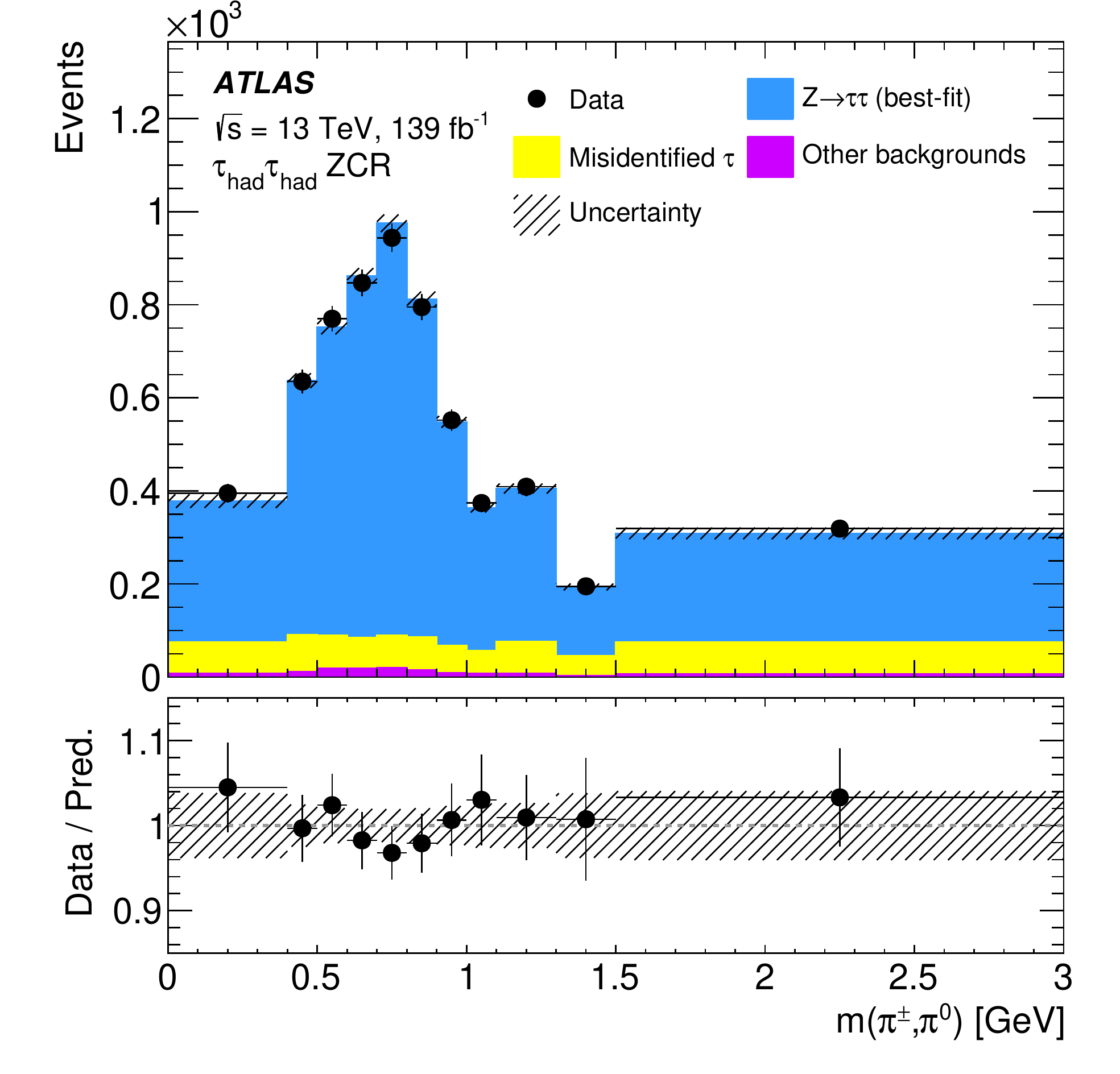}
\label{fig:ZttRhoCR_postfit_hh}
}
\caption{Post-fit distributions of the $\pi^{\pm}\pi^{0}$ invariant mass, $m(\pi^{\pm},\pi^{0})$,
in the $Z\to\tau\tau$ control regions in the
\protect\subref{fig:ZttRhoCR_postfit_lh} \tlhad and
\protect\subref{fig:ZttRhoCR_postfit_hh} \thadhad channels.
The $\ell$--1p1n (1p1n--1p1n) events are used in the \tlhad (\thadhad) channel in this control region.
For the \thadhad channel only the $m(\pi^{\pm},\pi^{0})$ value of the leading $\tauhad$ is selected.
\enquote{Other backgrounds} include $W$, diboson, top, $Z\to \ell\ell$ and $H\to WW^\ast$.
The hatched uncertainty band includes all sources of uncertainty after the fit to data.
}
\label{fig:ZttRhoCR_postfit}
\end{figure}

The uncertainties in the \pizero angular resolution and energy scale are estimated in situ in the analysis.
The \pizero energy and momentum direction are initially varied in accordance with the energy response of the calorimeter to charged pions~\cite{JETM-2018-08} and the measured \pizero angular resolution~\cite{PERF-2014-06}, respectively.
These variations in the \pizero energy scale and angular resolution are found to affect the shape of
the distribution of the reconstructed invariant mass of the $\pi^{\pm}\pi^{0}$ system, $m(\pi^{\pm}$, $\pi^{0})$,
in $\tau^{\pm}\rightarrow\rho^{\pm} \nu\rightarrow\pi^{\pm}\pi^{0}\nu$ decays.
Dedicated \Ztautau control regions with exclusive 1p1n decays, namely
the $\ell$--1p1n and 1p1n--1p1n events,
are defined in the \tlhad and \thadhad decay channels, respectively, as described in Section~\ref{sec:selection}.
The $m(\pi^{\pm}$, $\pi^{0})$ distribution is used as the observable in these control regions in the combined fit,
so that the final size of the \pizero angular resolution and energy scale uncertainties in the \phitau measurement
are determined from data.
Figure~\ref{fig:ZttRhoCR_postfit} shows the post-fit distributions of $m(\pi^{\pm},\pi^{0})$ in the $Z\to\tau\tau$ control regions in the \tlhad and \thadhad channels using
$\ell$--1p1n and 1p1n--1p1n events, respectively.
For the 1p1n--1p1n events in the \thadhad channel, only the leading $\tauhad$ is selected for the $m(\pi^{\pm},\pi^{0})$ distribution.
The $m(\pi^{\pm},\pi^{0})$ data distributions
show good agreement with the prediction.

% End of text imported from the .//source/systematics.tex input file

% The next lines are included from the .//source/statistical_analysis.tex input file
\section{Statistical analysis}
\label{sec:statistical_analysis}
 
To measure the $CP$-mixing angle \phitau, a simultaneous fit to the data is performed using a likelihood function
that depends on the $CP$-mixing angle \phitau as the parameter-of-interest, and the nuisance parameters that account for the systematic
uncertainties and the floating normalisations for the Higgs boson signal and for the background.
The likelihood function is constructed as a product of Poisson probability terms over the bins of the input distributions,
and the parameter-of-interest is estimated by maximising the likelihood.
The likelihood comprises 24 signal regions and 10 control regions.
Constraints on the nuisance parameters are assigned with a Gaussian term,
and bin-by-bin statistical fluctuations in the simulated background samples are included in the fit with a Poisson probability term.
 
In the fit the normalisation of the \Ztautau events is left to float freely in the eight control regions (described in Section~\ref{sec:selection})
to account for the \Ztautau modelling in the different signal regions.
Four normalisation factors (NF) are defined to control the \Ztautau background events in the Boost\_0, Boost\_1, VBF\_0, VBF\_1 categories, respectively.
Each factor is shared between the \tlhad and \thadhad channels, 
and is also shared between a control region and the corresponding signal region in the same category.
Other backgrounds are normalised to their
expected number of events estimated from the MC simulation.
 
Two control regions using events with a $\tau^{\pm}\rightarrow\rho^{\pm} \nu\rightarrow\pi^{\pm}\pi^{0}\nu$ decay ($\ell$--1p1n in \tlhad and 1p1n--1p1n in \thadhad) are defined from the \Ztautau control region events (Section~\ref{sec:selection}).
The distribution of the $\pi^{\pm}\pi^{0}$ invariant mass, $m(\pi^{\pm}$, $\pi^{0})$, is employed in these control regions to control the $\pi^0$-related uncertainties by using the data, as described in Section~\ref{sec:systematics}.
 
The Higgs boson signal strength $\mu_{\tau\tau}$ (defined as the ratio of the measured signal yield to the SM expectation) is also left unconstrained in the fit,
such that the signal normalisation does not depend on the SM assumption and only the shape of the \phistar distribution is exploited in the estimation of \phitau.
Model-dependence of the cross-section on $CP$-mixing scenarios is not exploited.
The \phistar distributions in the signal regions are binned to maximise the measurement sensitivity, taking into account the associated uncertainties.
In general, finer binnings are used in the signal regions associated with higher sensitivity (High and Medium SRs), with coarser binning in the Low signal region.
A smoothing procedure is applied in the signal regions to remove potentially large local fluctuations in
the systematic variations of the \phistar distributions because of the limited size of the MC samples used to build the templates.
The test statistic is based on a profile likelihood ratio and the asymptotic approximation~\cite{Cowan:2010js} is used for statistical interpretations.

% End of text imported from the .//source/statistical_analysis.tex input file

% The next lines are included from the .//source/results.tex input file
\FloatBarrier
 
\section{Results}
\label{sec:results}
 
\begin{figure}[t]
\centering
\subfloat[\tlhad High SR]{\includegraphics[height=0.26\textheight]{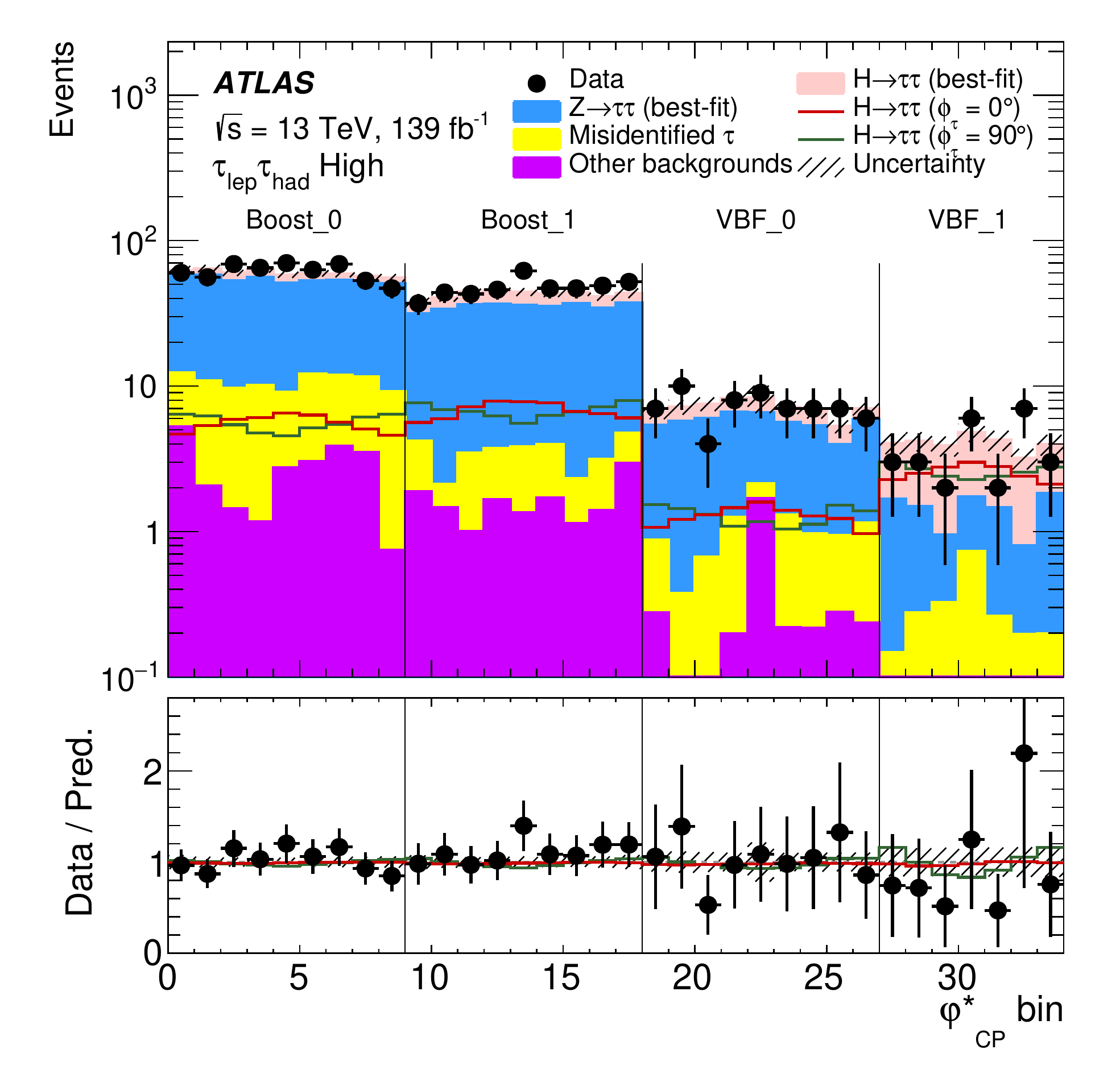}
\label{fig:phistar_high_postfit_lh}
}
\subfloat[\thadhad High SR]{\includegraphics[height=0.26\textheight]{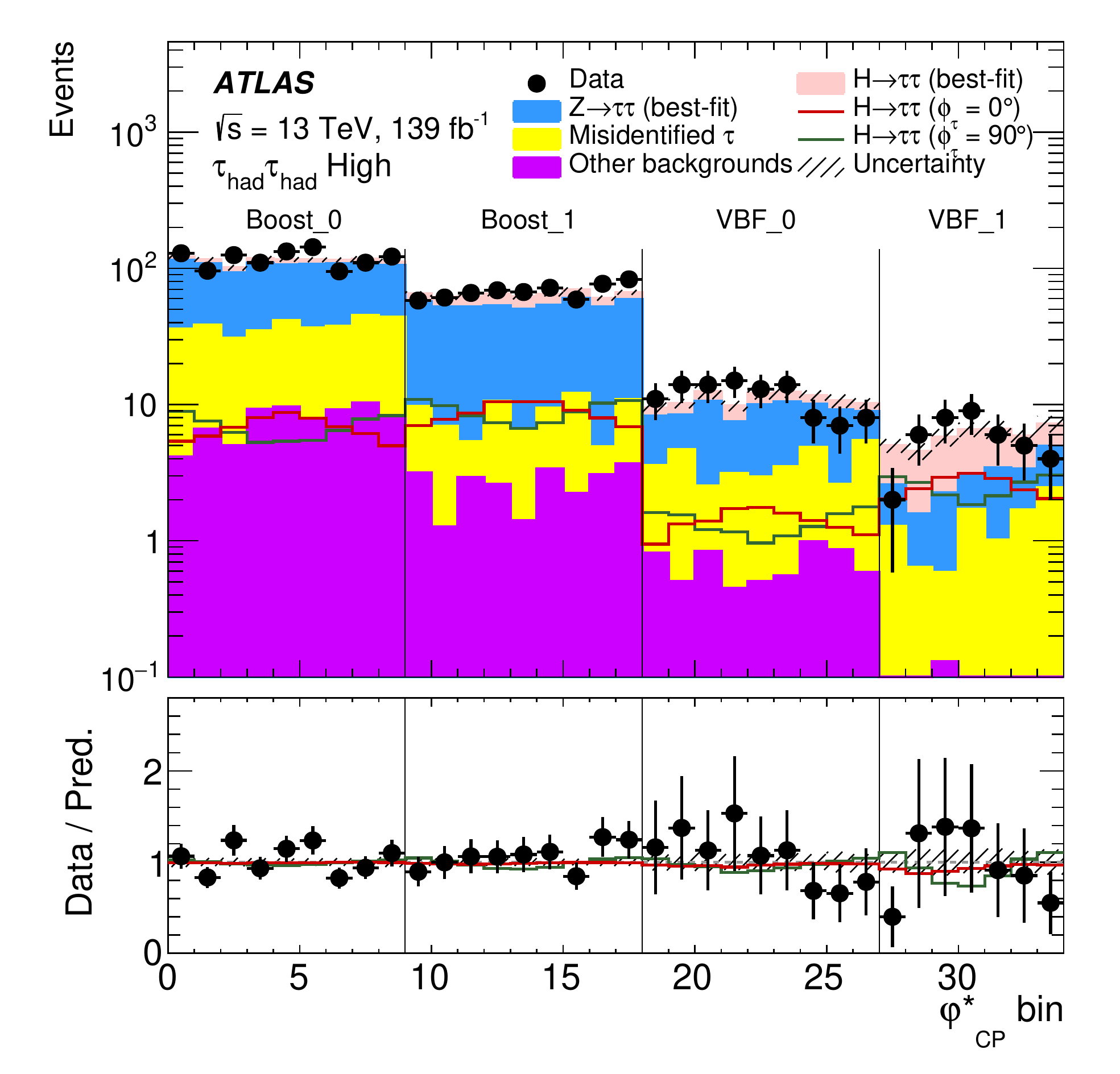}
\label{fig:phistar_high_postfit_hh}
}
\\
\subfloat[\tlhad Medium SR]{\includegraphics[height=0.26\textheight]{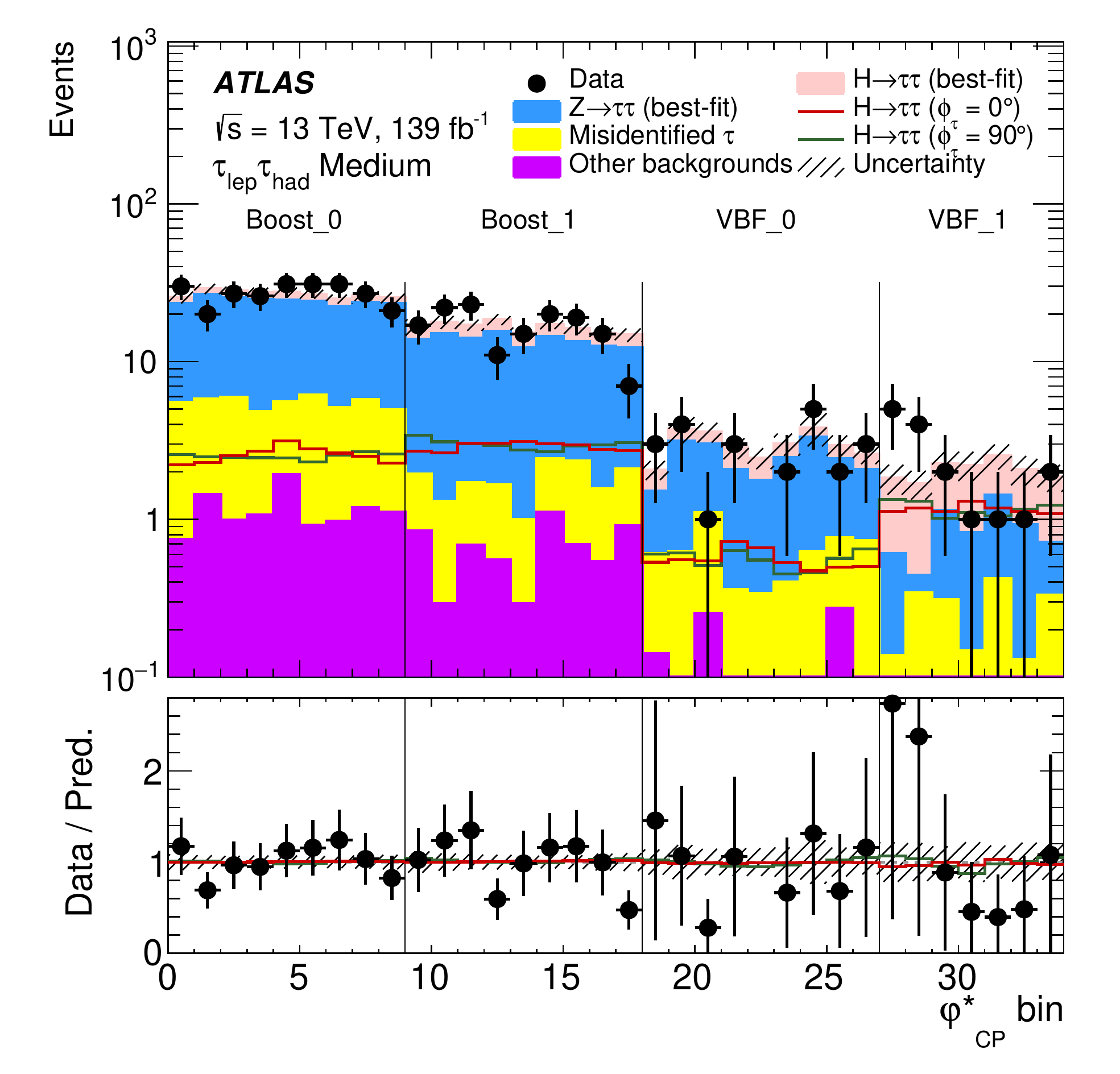}
\label{fig:phistar_medium_postfit_lh}
}
\subfloat[\thadhad Medium SR]{\includegraphics[height=0.26\textheight]{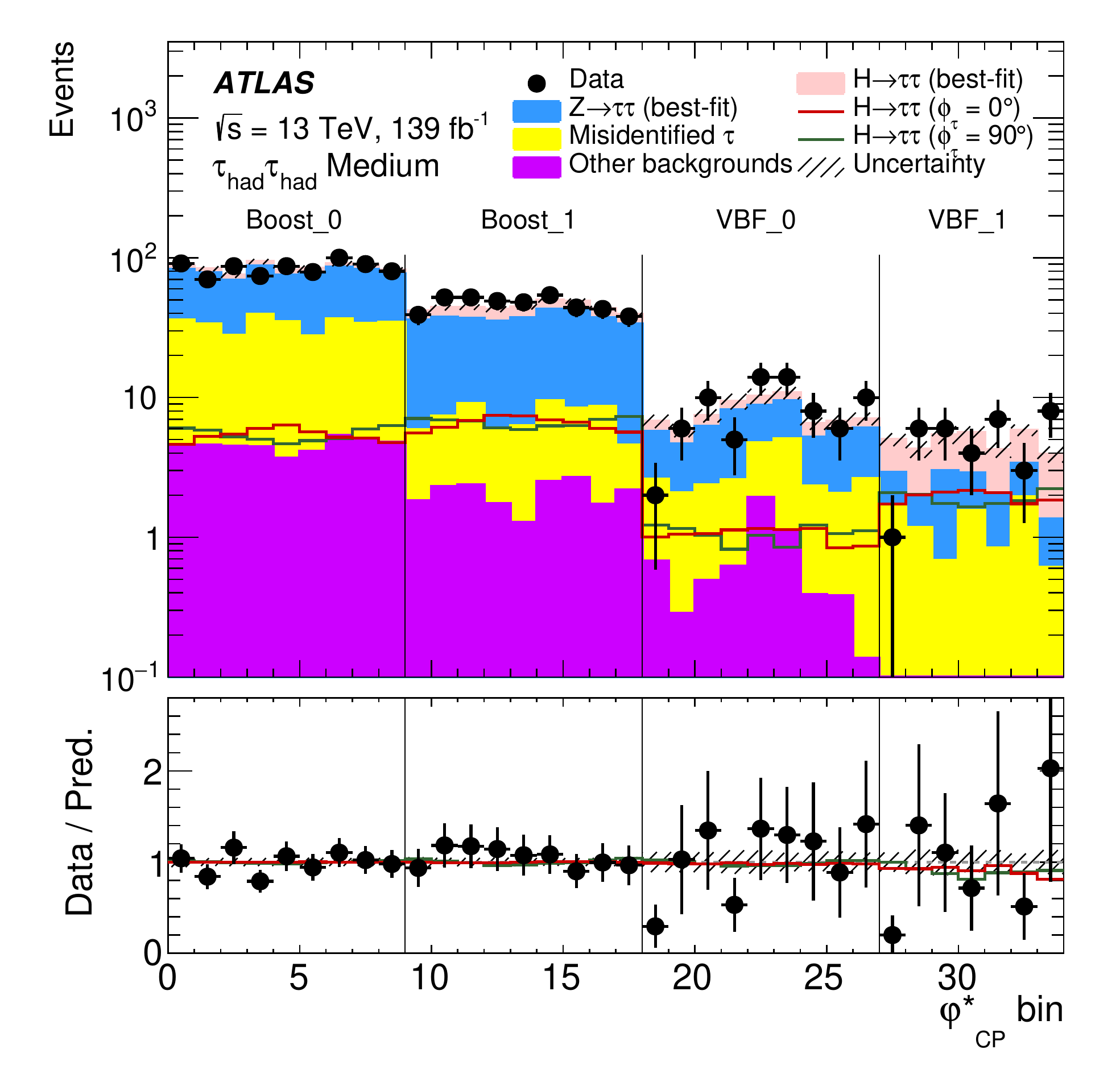}
\label{fig:phistar_medium_postfit_hh}
}
\\
\subfloat[\tlhad Low SR]{\includegraphics[height=0.26\textheight]{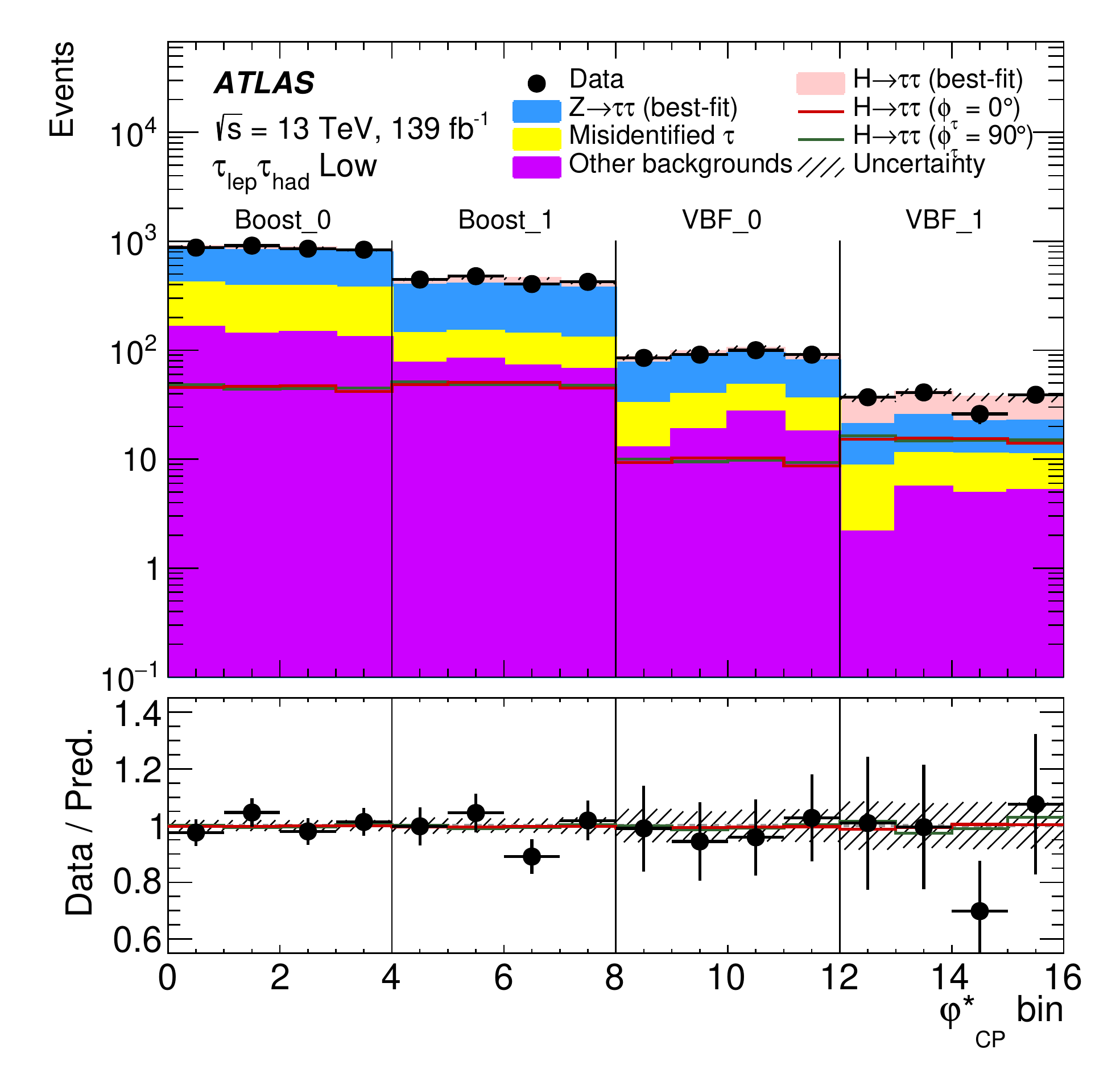}
\label{fig:phistar_low_postfit_lh}
}
\subfloat[\thadhad Low SR]{\includegraphics[height=0.26\textheight]{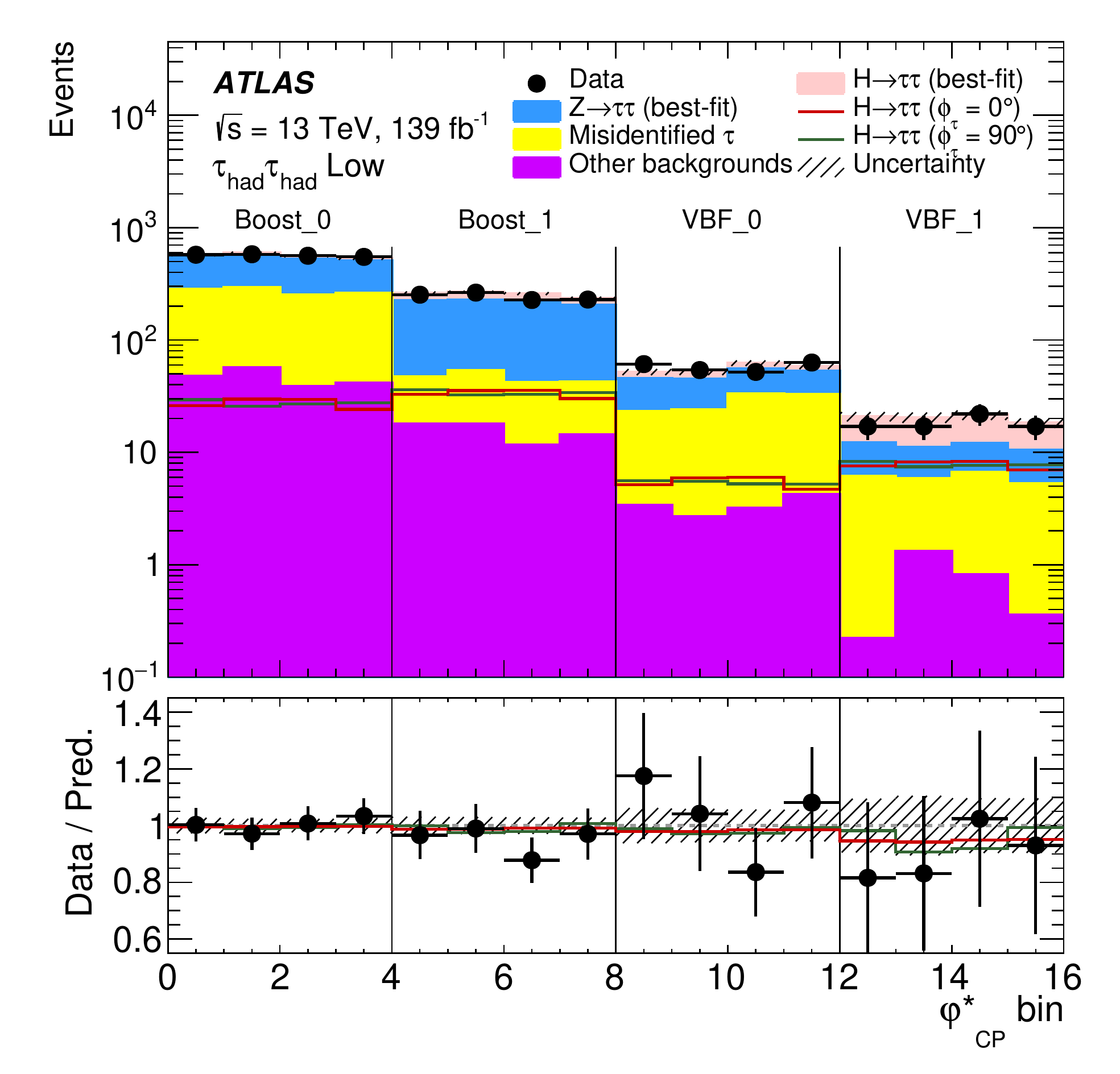}
\label{fig:phistar_low_postfit_hh}
}
\caption{Post-fit distributions of \phistar in the signal regions (SRs), showing
\protect\subref{fig:phistar_high_postfit_lh} \tlhad High SR,
\protect\subref{fig:phistar_high_postfit_hh} \thadhad High SR,
\protect\subref{fig:phistar_medium_postfit_lh} \tlhad Medium SR,
\protect\subref{fig:phistar_medium_postfit_hh} \thadhad Medium SR,
\protect\subref{fig:phistar_low_postfit_lh} \tlhad Low SR, and
\protect\subref{fig:phistar_low_postfit_hh} \thadhad Low SR.
The \phistar bins are counted incrementally through all VBF and Boost categories and cover the range [0, 360]$^{\circ}$ for each category.
The best-fit \Htautau signal is shown in solid pink,
while the red and green lines indicate the predictions for the pure $CP$-even (scalar, SM) and pure CP-odd (pseudoscalar) hypotheses, respectively,
scaled to the predicted signal yield.
\enquote{Other backgrounds} include $W$, diboson, top, $Z\to \ell\ell$ and $H\to WW^\ast$.
The hatched uncertainty band includes all sources of uncertainty after the fit to data.
}
\label{fig:phistar_SR_postfit}
\end{figure}

\begin{figure}[t]
\centering
\includegraphics[height=0.4\textheight]{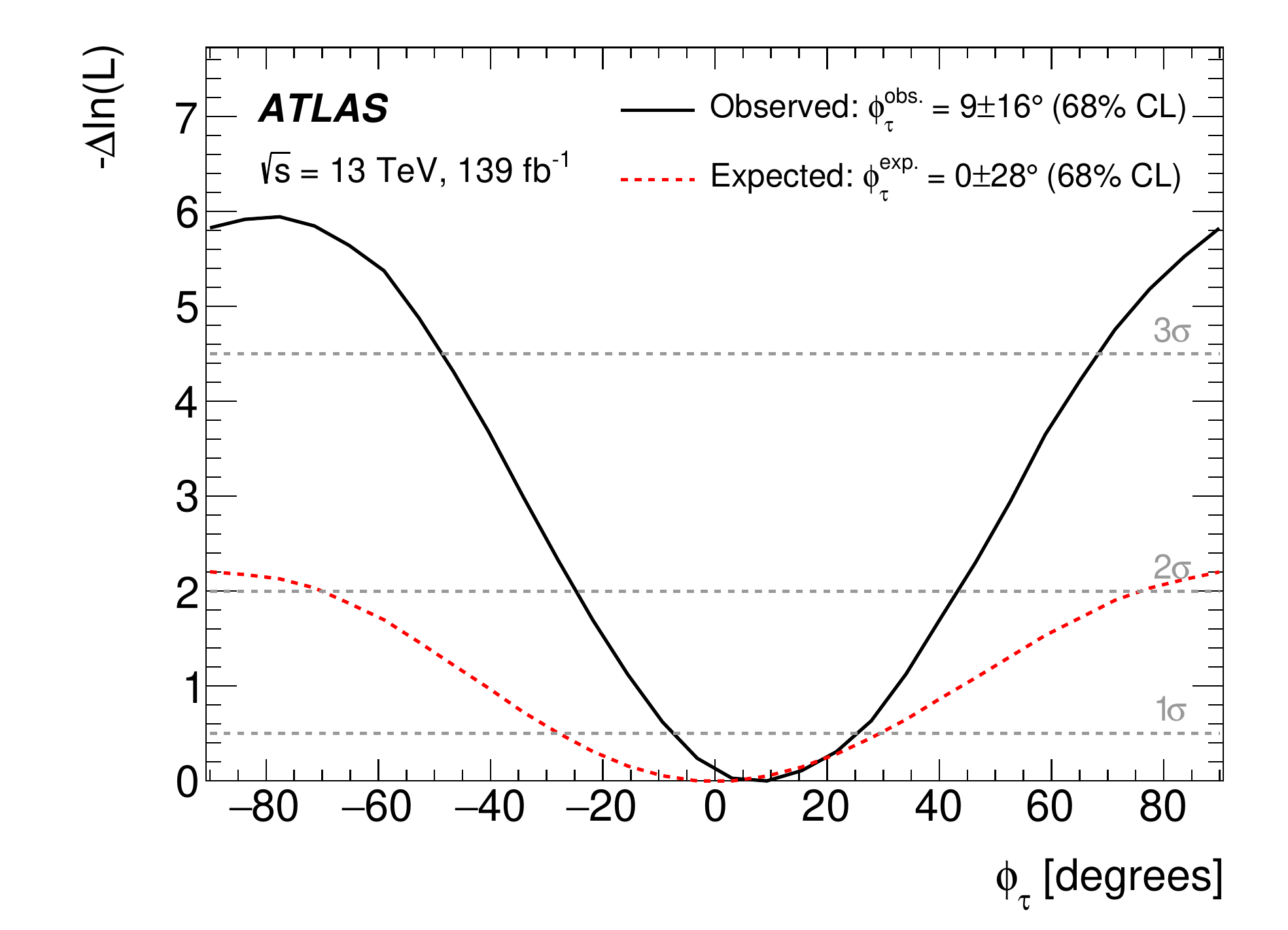}
\caption{One-dimensional likelihood scan of the $CP$-mixing angle \phitau.
The observed (expected) value of $\phi_{\tau}$ is $9^{\circ}\pm16^{\circ}$ ($0^{\circ}\pm28^{\circ}$) at the 68\% confidence level (CL), and $\pm 34^{\circ}$ ($_{-70^{\circ}}^{+75^{\circ}}$) at the 2$\sigma$ level.
The $CP$-odd hypothesis is rejected at the 3.4$\sigma$ (2.1$\sigma$ expected) level.}
\label{fig:1DLH}
\end{figure}

Figure~\ref{fig:phistar_SR_postfit} shows the post-fit \phistar distributions for the data as well as the prediction in the High, Medium and Low signal regions in the \tlhad and \thadhad channels, respectively.
In each distribution, the \phistar bins are counted incrementally through all VBF and Boost categories and cover the range [0, 360]$^{\circ}$ for each category.
The observed and expected negative log-likelihood ($\Delta\ln L$)
scans in \phitau are shown in Figure~\ref{fig:1DLH}.
The observed (expected) value of $\phi_{\tau}$ is
$9^{\circ} \pm 16^{\circ}$ ($0^{\circ} \pm 28 ^{\circ}$)
at the 68\% confidence level (CL),
and
$\pm 34^{\circ}$ ($_{-70^{\circ}}^{+75^{\circ}}$)
at the 2$\sigma$ level.
The data disfavours the pure $CP$-odd hypothesis at the 3.4$\sigma$ level, while the expected exclusion level is 2.1$\sigma$.
The fitted signal and background normalisations are shown in Table~\ref{tab:postfit_parameters}.
The results are compatible with the SM expectation within the measured uncertainties.

The total uncertainty is dominated by the statistical uncertainties of the data sample.
The dominant contributions to the systematic uncertainties are from jets,
followed by limited sample size of the background simulations,
uncertainties from the free-floating normalisation factors,
and theory uncertainties.
The uncertainties in the $\tau$-lepton decay reconstruction have small impacts of less than 1$^{\circ}$ on \phitau.
The effects from the $\pi^{0}$ uncertainties measured in situ are compared with those from a set of simulated event samples with systematically varied modelling of the hadronic response in the calorimeter, and the latter are found to have a similar impact on the \phitau measurement.
Effects from other sources are negligible.
The impact of the uncertainties is summarised in Table~\ref{tab:uncertainty_impact}.

The difference between the observed and expected sensitivities of the \phitau measurement can be attributed to a statistical fluctuation in data.
The uncertainties of the \phitau measurement are highly dependent on the size of the modulation amplitude of the \phistar distributions.
In the measured data, there are more distortions of the \phistar shape due to bins fluctuating from their expected values, resulting in  overall larger modulation amplitudes than the predicted shape.
A set of pseudo-experiments is performed (with the condition $\mu_{\tau\tau} = 1$)
that shows that the probability of obtaining the observed distribution,
given the expectation, is about 4\%.

The present measurement is compatible with the recent measurement of the same mixing-angle parameter by the CMS Collaboration~\cite{CMS-HIG-20-006}, for which an observed (expected) mixing-angle value of $-1^{\circ}\pm19^{\circ}$ ($0^{\circ}\pm21^{\circ}$) at the 68\% CL was reported.
 
The expected sensitivities of the  \thadhad and \tlhad channels in excluding a pure $CP$-odd $H\tau\tau$ coupling are 1.7$\sigma$ and 1.1$\sigma$, respectively.
The `High' signal regions contribute the most, with 1.4$\sigma$ and 1.0$\sigma$ in the \thadhad and \tlhad channels, respectively.
Other signal regions have expected sensitivities below 1$\sigma$.
 
\begin{figure}[t]
\centering
\includegraphics[height=0.4\textheight]{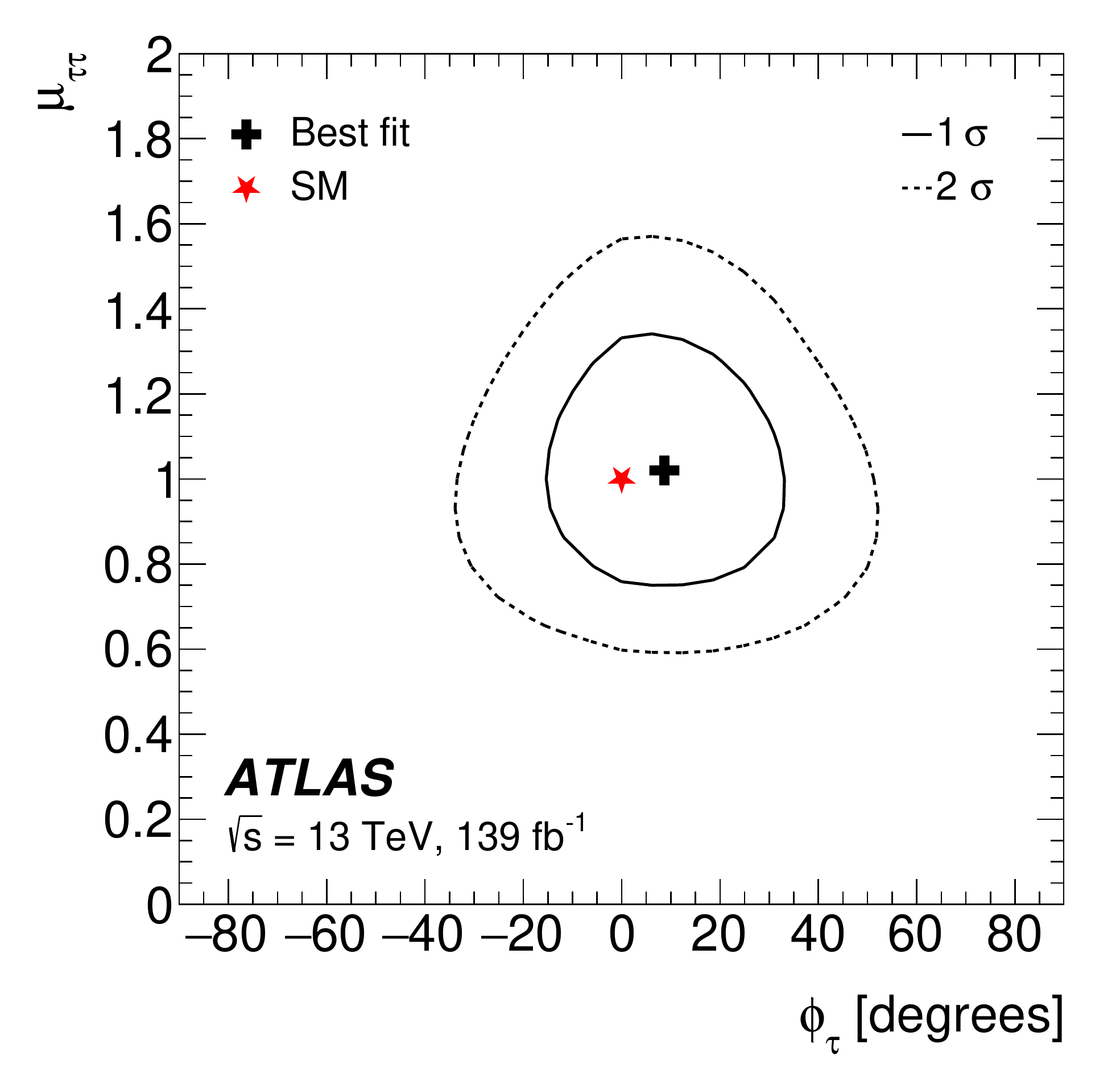}
\caption{A 2D likelihood scan of the observed signal strength $\mu_{\tau\tau}$ versus the $CP$-mixing angle \phitau.
The 1$\sigma$ and 2$\sigma$ confidence regions are shown.
}
\label{fig:2DLH}
\end{figure}

A 2D scan of $\Delta\ln L$ as a function of the signal strength $\mu_{\tau\tau}$ versus \phitau is shown in Figure~\ref{fig:2DLH}.
The 1$\sigma$ and 2$\sigma$ 2D confidence levels correspond to $\Delta\ln L$ values of 1.15 and 3.09,
respectively. 
No strong correlation between $\mu_{\tau\tau}$ and \phitau is observed.
The SM prediction of $\mu_{\tau\tau} = 1$ and $\phitau = 0$ is compatible with the measurement within the 1$\sigma$ confidence region.

\begin{figure}[t]
\centering
\includegraphics[height=0.4\textheight]{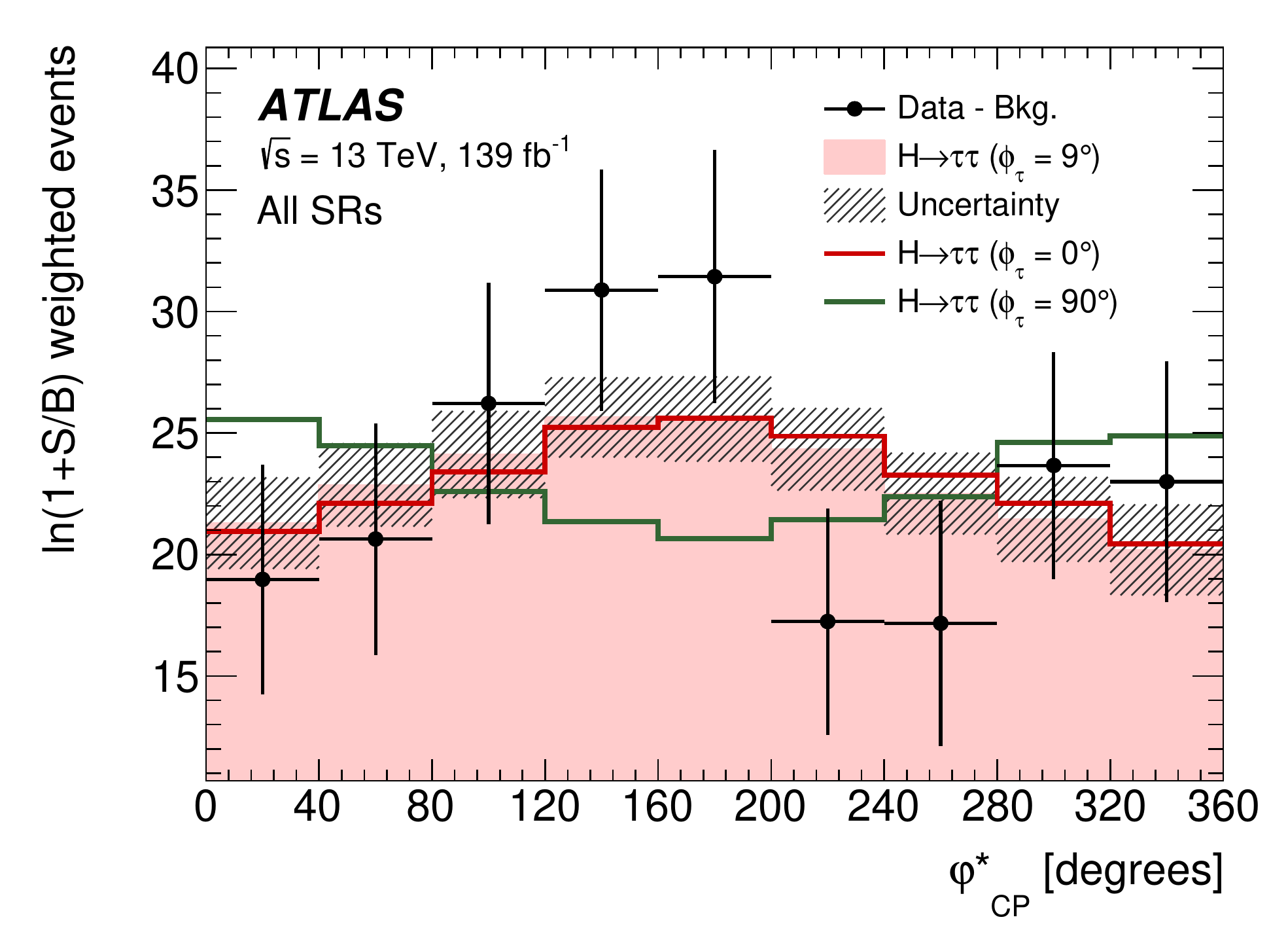}
\caption{
Combined post-fit distribution of \phistar from all signal regions in both the \tlhad and \thadhad channels.
Events are weighted with $\ln(1+S/B)$ for the corresponding signal region.
The background is subtracted from data.
The best-fit \Htautau signal is shown in solid pink, while the red and green lines indicate the predictions for the pure $CP$-even (scalar, SM) and pure CP-odd (pseudoscalar) hypotheses, respectively,
all scaled to the best-fit \Htautau signal yield.
The hatched uncertainty band includes all sources of uncertainty after the fit to data, and represents the same uncertainty in the total signal and background predictions as in
Figure~\ref{fig:phistar_SR_postfit}.
}
\label{fig:data_minus_bkg}
\end{figure}

Figure~\ref{fig:data_minus_bkg} shows the data distribution of \phistar with all signal regions in both the \tlhad and \thadhad channels combined, together with the best-fit \Htautau signal, pure $CP$-even and pure $CP$-odd hypotheses. The events in each signal region are weighted with $\ln(1+S/B)$, where $S$ and $B$ are the event yields for the signal and the total background, respectively.
The background is subtracted from data in the figure.
The distribution illustrates that the data disfavours the pure $CP$-odd scenario.

\FloatBarrier
 
\begin{table}[t]
\renewcommand{\arraystretch}{1.4}
\centering
\caption{Free-floating parameters in the measurement. Observed and expected values are shown for the $CP$-mixing angle (\phitau), the signal strength ($\mu_{\tau\tau}$) and various background normalisations for \Ztautau (NF$_{\Ztautau}$)
corresponding to different signal phase-space regions.
}
\begin{tabular}{lcc}
\toprule
Fitted parameters             & Observed & Expected \\
\midrule
\phitau               & $9^{\circ}\pm16^{\circ}$ & $0^{\circ}\pm28^{\circ}$ \\
\midrule
$\mu_{\tau\tau}$              & $1.02^{+0.20}_{-0.20}$ & $1.00^{+0.21}_{-0.21}$ \\
\midrule
NF$_{\Ztautau}^{\mathrm{Boost\_1}}$           & $1.01\pm0.05$ & $1.00\pm0.04$ \\
NF$_{\Ztautau}^{\mathrm{Boost\_0}}$           & $1.02\pm0.05$ & $1.00\pm0.05$ \\
NF$_{\Ztautau}^{\mathrm{VBF\_1}}$             & $1.04\pm0.08$ & $1.00\pm0.08$ \\
NF$_{\Ztautau}^{\mathrm{VBF\_0}}$             & $0.95\pm0.07$ & $1.00\pm0.08$ \\
\bottomrule
\end{tabular}
\label{tab:postfit_parameters}
\end{table}
 
\FloatBarrier
 
\begin{table}[t]
\centering
\caption{Impact of different sources of uncertainty on the \phitau measurement.
}
\begin{tabular}{lc}
\toprule
Set of nuisance parameters   &  Impact on \phitau[degrees] \\
\midrule
Jet energy scale                                  & 3.4 \\ 
Jet energy resolution                                  & 2.5 \\ 
Pile-up jet tagging                                & 0.5 \\ 
Jet flavour tagging                       & 0.2 \\ 
\met                                  & 0.4 \\ 
Electron                              & 0.3 \\ 
Muon                                  & 0.9 \\ 
\tauhad reconstruction                 & 1.0 \\ 
Misidentified $\tau$                  & 0.6 \\ 
\tauhad decay mode classification                 & 0.3 \\ 
\pizero angular resolution and energy scale     & 0.2 \\ 
Track ($\pi^{\pm}$, impact parameter)  & 0.7 \\ 
Luminosity                            & 0.1 \\ 
Theory uncertainty in $H\to\tau\tau$ processes  & 1.5 \\ 
Theory uncertainty in $Z\to\tau\tau$ processes  & 1.1 \\ 
Simulated background sample statistics         & 1.4 \\ 
Signal normalisation   & 1.4 \\ 
Background normalisation   & 0.6 \\ 
\midrule
Total systematic uncertainty          & 5.2 \\ 
Data sample statistics                  & 15.6\phantom{0} \\ 
\midrule
Total                                & 16.4\phantom{0} \\ 
\bottomrule
\end{tabular}
\label{tab:uncertainty_impact}
\end{table}
 
\FloatBarrier
 
\FloatBarrier

% End of text imported from the .//source/results.tex input file

% The next lines are included from the .//source/conclusion.tex input file
\section{Conclusion}
\label{sec:conclusion}
 
A measurement of the $CP$ properties of the interaction between the Higgs boson and $\tau$-leptons is presented.
In the generalised Yukawa interaction, a single mixing parameter \phitau  parameterises $CP$-violating interactions between the Higgs boson and $\tau$-leptons.
The measurement is performed using  139~fb$^{-1}$ of proton--proton collision data recorded at a centre-of-mass
energy of $\sqrt{s} = 13$~\TeV\ with the ATLAS detector at the Large Hadron Collider.
The measurement is based on a maximum-likelihood fit to the $CP$-sensitive angular observable \phistar defined by the
visible decay products in the \tlhad and \thadhad decay channels.
Depending on the decay channels, different methods are used to reconstruct the
\phistar distributions. The sensitivity is optimised by applying decay-mode-dependent kinematic selections.
The observed (expected) value of \phitau is $9^{\circ} \pm 16^{\circ}$ ($0^{\circ} \pm 28^{\circ}$)
at the 68\% confidence level, 
and
$\pm 34^{\circ}$ ($_{-70^{\circ}}^{+75^{\circ}}$)
at the 2$\sigma$ level.
The pure $CP$-odd hypothesis is disfavoured at a level of 3.4~standard deviations.
The analysis precision is limited by the statistical uncertainty of the data sample.
The measurement is consistent with the Standard Model expectation.

% End of text imported from the .//source/conclusion.tex input file

\FloatBarrier

\section*{Acknowledgements}

% The next lines are included from the .//acknowledgements/Acknowledgements.tex input file

We thank CERN for the very successful operation of the LHC, as well as the
support staff from our institutions without whom ATLAS could not be
operated efficiently.
 
We acknowledge the support of
ANPCyT, Argentina;
YerPhI, Armenia;
ARC, Australia;
BMWFW and FWF, Austria;
ANAS, Azerbaijan;
CNPq and FAPESP, Brazil;
NSERC, NRC and CFI, Canada;
CERN;
ANID, Chile;
CAS, MOST and NSFC, China;
Minciencias, Colombia;
MEYS CR, Czech Republic;
DNRF and DNSRC, Denmark;
IN2P3-CNRS and CEA-DRF/IRFU, France;
SRNSFG, Georgia;
BMBF, HGF and MPG, Germany;
GSRI, Greece;
RGC and Hong Kong SAR, China;
ISF and Benoziyo Center, Israel;
INFN, Italy;
MEXT and JSPS, Japan;
CNRST, Morocco;
NWO, Netherlands;
RCN, Norway;
MEiN, Poland;
FCT, Portugal;
MNE/IFA, Romania;
MESTD, Serbia;
MSSR, Slovakia;
ARRS and MIZ\v{S}, Slovenia;
DSI/NRF, South Africa;
MICINN, Spain;
SRC and Wallenberg Foundation, Sweden;
SERI, SNSF and Cantons of Bern and Geneva, Switzerland;
MOST, Taiwan;
TENMAK, T\"urkiye;
STFC, United Kingdom;
DOE and NSF, United States of America.
In addition, individual groups and members have received support from
BCKDF, CANARIE, Compute Canada and CRC, Canada;
PRIMUS 21/SCI/017 and UNCE SCI/013, Czech Republic;
COST, ERC, ERDF, Horizon 2020 and Marie Sk{\l}odowska-Curie Actions, European Union;
Investissements d'Avenir Labex, Investissements d'Avenir Idex and ANR, France;
DFG and AvH Foundation, Germany;
Herakleitos, Thales and Aristeia programmes co-financed by EU-ESF and the Greek NSRF, Greece;
BSF-NSF and MINERVA, Israel;
Norwegian Financial Mechanism 2014-2021, Norway;
NCN and NAWA, Poland;
La Caixa Banking Foundation, CERCA Programme Generalitat de Catalunya and PROMETEO and GenT Programmes Generalitat Valenciana, Spain;
G\"{o}ran Gustafssons Stiftelse, Sweden;
The Royal Society and Leverhulme Trust, United Kingdom.
 
The crucial computing support from all WLCG partners is acknowledged gratefully, in particular from CERN, the ATLAS Tier-1 facilities at TRIUMF (Canada), NDGF (Denmark, Norway, Sweden), CC-IN2P3 (France), KIT/GridKA (Germany), INFN-CNAF (Italy), NL-T1 (Netherlands), PIC (Spain), ASGC (Taiwan), RAL (UK) and BNL (USA), the Tier-2 facilities worldwide and large non-WLCG resource providers. Major contributors of computing resources are listed in Ref.~\cite{ATL-SOFT-PUB-2021-003}.

% End of text imported from the .//acknowledgements/Acknowledgements.tex input file

\clearpage

\printbibliography

\clearpage
 
% ATLAS Collaboration author list
% Reference date of HIGG-2019-10 is 2022-04-26
% Author list last updated on date 07-JUL-23
% Data extracted on 07-Jul-2023 for paper reference HIGG-2019-10
% at 11:23am
 
\begin{flushleft}
\hypersetup{urlcolor=black}
{\Large The ATLAS Collaboration}

\bigskip

\AtlasOrcid[0000-0002-6665-4934]{G.~Aad}$^\textrm{\scriptsize 101}$,
\AtlasOrcid[0000-0002-5888-2734]{B.~Abbott}$^\textrm{\scriptsize 119}$,
\AtlasOrcid[0000-0002-7248-3203]{D.C.~Abbott}$^\textrm{\scriptsize 102}$,
\AtlasOrcid[0000-0002-1002-1652]{K.~Abeling}$^\textrm{\scriptsize 55}$,
\AtlasOrcid[0000-0002-8496-9294]{S.H.~Abidi}$^\textrm{\scriptsize 29}$,
\AtlasOrcid[0000-0002-9987-2292]{A.~Aboulhorma}$^\textrm{\scriptsize 35e}$,
\AtlasOrcid[0000-0001-5329-6640]{H.~Abramowicz}$^\textrm{\scriptsize 150}$,
\AtlasOrcid[0000-0002-1599-2896]{H.~Abreu}$^\textrm{\scriptsize 149}$,
\AtlasOrcid[0000-0003-0403-3697]{Y.~Abulaiti}$^\textrm{\scriptsize 116}$,
\AtlasOrcid[0000-0003-0762-7204]{A.C.~Abusleme~Hoffman}$^\textrm{\scriptsize 136a}$
% ...
\bigskip
\\

% ...
$^{29}$Physics Department, Brookhaven National Laboratory, Upton NY; United States of America.\\
% ...
$^{35}$$^{(a)}$Facult\'e des Sciences Ain Chock, R\'eseau Universitaire de Physique des Hautes Energies - Universit\'e Hassan II, Casablanca;$^{(b)}$Facult\'{e} des Sciences, Universit\'{e} Ibn-Tofail, K\'{e}nitra;$^{(c)}$Facult\'e des Sciences Semlalia, Universit\'e Cadi Ayyad, LPHEA-Marrakech;$^{(d)}$LPMR, Facult\'e des Sciences, Universit\'e Mohamed Premier, Oujda;$^{(e)}$Facult\'e des sciences, Universit\'e Mohammed V, Rabat;$^{(f)}$Institute of Applied Physics, Mohammed VI Polytechnic University, Ben Guerir; Morocco.\\
% ...
$^{55}$II. Physikalisches Institut, Georg-August-Universit\"{a}t G\"ottingen, G\"ottingen; Germany.\\
% ...
$^{101}$CPPM, Aix-Marseille Universit\'e, CNRS/IN2P3, Marseille; France.\\
$^{102}$Department of Physics, University of Massachusetts, Amherst MA; United States of America.\\
% ...
$^{116}$Department of Physics, New York University, New York NY; United States of America.\\
% ...
$^{119}$Homer L. Dodge Department of Physics and Astronomy, University of Oklahoma, Norman OK; United States of America.\\
% ...
$^{136}$$^{(a)}$Departamento de F\'isica, Pontificia Universidad Cat\'olica de Chile, Santiago;$^{(b)}$Millennium Institute for Subatomic physics at high energy frontier (SAPHIR), Santiago;$^{(c)}$Instituto de Investigaci\'on Multidisciplinario en Ciencia y Tecnolog\'ia, y Departamento de F\'isica, Universidad de La Serena;$^{(d)}$Universidad Andres Bello, Department of Physics, Santiago;$^{(e)}$Instituto de Alta Investigaci\'on, Universidad de Tarapac\'a, Arica;$^{(f)}$Departamento de F\'isica, Universidad T\'ecnica Federico Santa Mar\'ia, Valpara\'iso; Chile.\\
% ...
$^{149}$Department of Physics, Technion, Israel Institute of Technology, Haifa; Israel.\\
$^{150}$Raymond and Beverly Sackler School of Physics and Astronomy, Tel Aviv University, Tel Aviv; Israel.\\
% ...
$^{a}$ Also Affiliated with an institute covered by a cooperation agreement with CERN.\\
$^{b}$ Also at An-Najah National University, Nablus; Palestine.\\
$^{c}$ Also at Borough of Manhattan Community College, City University of New York, New York NY; United States of America.\\
$^{d}$ Also at Bruno Kessler Foundation, Trento; Italy.\\
$^{e}$ Also at Center for High Energy Physics, Peking University; China.\\
$^{f}$ Also at Centro Studi e Ricerche Enrico Fermi; Italy.\\
% ...

\end{flushleft}

% Created with Glance <Atlas.Glance@cern.ch>

\end{document}